\documentclass[12pt]{article}
\usepackage{graphicx, amsmath, tikz, graphicx, cite, color, braket, caption, subcaption} 
\usepackage[margin = 2cm]{geometry}
\usepackage[colorlinks=true]{hyperref}

\tikzset{every loop/.style={min distance=15mm,looseness=10}}

\DeclareMathOperator{\nbd}{nbd}
\DeclareMathOperator{\tr}{Tr}

\title{Non-Markovianity in Discrete-Time Open Quantum Random Walk on Arbitrary Graphs}
\author{
	Monika Rani$^1$\thanks{Email: \texttt{rani.2@iitj.ac.in}}, Supriyo Dutta$^2$\thanks{Email: \texttt{dosupriyo@gmail.com}}, Subhashish Banerjee$^1$\thanks{Email: \texttt{subhashish@iitj.ac.in}} \\
	\small{$^1$Department of Physics, Indian Institute of Technology Jodhpur} \\
	\small{Jodhpur, Rajasthan, India-342037.} \vspace{.15cm}\\
	\small{$^2$Department of Mathematics, National Institute of Technology Agartala} \\
	\small{Jirania, West Tripura, India - 799046.}
}
\date{} 

\begin{document}
	
	\maketitle 
	
	\begin{abstract}
		In this work, we present a new model of the Discrete-Time Open Quantum Walk (DTOQW) applicable to an arbitrary graph, thereby going beyond the case of quantum walks on regular graphs. We study the impact of noise in the dynamics of quantum walk by applying Kraus operators of different dimensions which are constructed using the Weyl operators. The DTOQW employs these Kraus operators as its coin operators. The walker dynamics are studied under the impact of non-Markovian amplitude damping, dephasing and depolarizing noise channels. We also implement the walk on various graphs, including path graphs, cycle graphs, star graphs, complete graphs, complete bipartite graphs, etc. We gauge the dynamics by computing coherence and fidelity at different time steps, taking into account the influence of noise. Furthermore, we compute the probability distribution at different time steps for the above noises, which represents the availability of the quantum walker at different vertices of the graph. 
	\end{abstract}

	\section{Introduction}
	
	The idea of quantum walk \cite{kempe2003quantum, konno2008quantum, venegas2012quantum, portugal2013quantum}, a quantum mechanical analogue of the classical random walk \cite{klafter2011first}, is a crucial tool for in quantum computation and information \cite{nielsen2010quantum, mcmahon2007quantum}. Different variants of quantum walks are applicable in quantum algorithms \cite{venegas2012quantum, xia2019random, motta2022emerging}, quantum simulation \cite{childs2009universal}, quantum communication \cite{chen2019quantum, mukai2020discrete}, quantum state transfer \cite{vstefavnak2017perfect, li2021discrete, dutta2022perfect}, quantum search algorithms \cite{ambainis2020quadratic}, element distinctness \cite{ambainis2007quantum, buhrman2001quantum} quantum routing \cite{li2021discrete, shi2023quantum, dutta2023quantum}, ranking the vertices in a network \cite{paparo2012google, dutta2024discrete}, among others. Different models of quantum walks include multiple variants of discrete-time quantum walk \cite{watrous2001quantum, aharonov2001quantum}, continuous-time quantum walk \cite{farhi1998quantum, godsil2008periodic}, Szegedy quantum walk \cite{szegedy2004quantum, szegedy2004spectra}, Fibonacci quantum walk \cite{romanelli2009fibonacci}, open quantum walk \cite{attal2012open}. The original ideas of discrete-time quantum walk, continuous-time quantum walk, Szegedy quantum walk and Fibonacci quantum walk are governed by a unitary operator that has no interaction with an external environment. Studying quantum walks in noisy environments is essential because real-world quantum systems are inevitably subject to various forms of noise \cite{banerjee2018open}. It significantly affects the behaviour and performance of quantum walk \cite{kendon2007decoherence, shapira2003one}. Understanding the effect of these noises is crucial for designing robust quantum systems. The experimental implementations of quantum walk also analyze the effects of noise \cite{chandrashekar2007symmetries, romanelli2005decoherence, zahringer2010realization, yin2008quantum, broome2010discrete}.
	
	There are multiple proposals to incorporate the environmental noise in quantum walk \cite{chandrashekar2007symmetries, banerjee2008symmetry, chandrashekar2010relationship, rao2011quantumness}. A number of quantum operators are used as coin operators in discrete-time quantum walks, such as the Hadamard operator and Grover operator. Generally, in traditional quantum walk models, noise and coin operators are treated separately \cite{chandrashekar2006discrete, wong2015grover, kumar2018non,attal2012open,banerjee2017non,kumar2018enhanced}. Noise is usually applied as a separate process, mostly after the coin operation. The effects of noise on the continuous-time quantum walk have also been discussed in the literature \cite{benedetti2016non, rossi2017continuous}. 
	
	The open quantum walk \cite{attal2012open, wang2018open, dhahri2019open, liu2017steady, mukhamedov2022open, carbone2015homogeneous, kemp2020lazy} is a model of quantum walk which offers a platform to investigate the effect of noise on the dynamics of the quantum walk. The Discrete-Time Open Quantum Walk (DTOQW) is the discrete variant of an open quantum walk. In this work, we present a new model of DTOQW in a noisy environment on an arbitrary graph. In contrast to the other models of quantum walk in a noisy environment, our model integrates the noise directly into the coin operators, making the noise an intrinsic part of the quantum walk dynamics.  Understanding the properties of the associated noisy quantum channel is necessary to explore the effect of noise. We utilize a number of noisy quantum channels for our investigation, such as the non-Markovian Amplitude Damping Channel (ADC), non-Markovain dephasing (NMD) noise and non-Markovian depolarizing noise \cite{dutta2023qudit, shrikant2018non, han2024construction}.
	
	The new model of DTOQW is defined on an arbitrary graph. This is another advantage of the new model, as the previous models of DTOQW were defined on regular graphs, where the degree of each vertex is the same, such as the infinite path graph, cycle graph \cite{attal2012open, sinayskiy2019open, sinayskiy2013open}. In this article, we break this limitation. In our model, the vertices of the graph provide the location of the walker. The possible directions in which the walker can move are represented by the directed edges in the graph. 
	
	To study the effect of noise we consider a number of quantum channels which were originally defined for qubit states. We can generalize these channels for qudits of arbitrary dimension using the Weyl operators \cite{dutta2023qudit, basile2024weyl}. This is essential as the degree of the vertices in an arbitrary graph is not unique. Corresponding to every outgoing edge from a vertex we assign a Kraus operator representing a noisy quantum channel. Our analysis involves the calculation of fidelity between the initial and final states of the walker at different time steps. It provides a quantitative measure of the impact of noise on quantum walker dynamics. Also, we work out the probability distribution for each vertex in different time steps. Coherence and fidelity are observed as a function of graph steps for various types of noise.
	
	This paper is organized as follows. The preliminary ideas are elucidated in section 2. In section 3, we construct the coin operator of DTOQW on an arbitrary graph.  Here, we also generalize different noisy channels such as the non-Markovian amplitude damping channel, non-Markovian dephasing channel and non-Markovian depolarization channel, for arbitrary dimensions. In Section 4, we construct our DTOQW model on arbitrary graphs. In different subsections, we discuss DTOQW on different types of graphs, such as the path graph, cycle graph, star graph, complete graph, etc. Also, we consider a graph which does not belong to any of these classes in support to our claim that this new model of quantum walk is applicable for all graphs. The probability distribution of the walker at different vertices of the graph, fidelity between the initial and final state as well as the coherence of the quantum state of the walker are calculated for different graphs. We then conclude this article. In the conclusion, we illustrate some applications of the new quantum walk model. The completeness conditions for Kraus operators considered in this work are relegated to the Appendix.

	\section{Preliminaries}
	
	A graph $G = (V(G), E(G))$ is a triple consisting of a vertex set $V(G)$, a set of edges $E(G)$ and a relationship that associates two vertices with each edge \cite{west2001introduction, bondy1976graph}. The number of vertices in a graph is called the order of the graph, which is denoted by $n$, throughout this article. A vertex $u$ is said to be adjacent to a vertex $v$ if $(u, v) \in E(G)$. Also, the edge $(u, v)$ is incident to $u$ and $v$. A loop is an edge of the form $(u, u)$ that connects a vertex $u$ with itself. The neighbourhood of $u$ is denoted by $\nbd(u)$ which is the set of all vertices adjacent to it. The number of elements in $\nbd(u)$ is the degree of vertex $u$, which we denote as $d_u$. We draw a vertex by a marked circle and an edge by a straight line joining two circles. 
	
	An edge with a direction is called a directed edge. A directed edge from the vertex $u$ to the vertex $v$ is denoted by $\overrightarrow{(u, v)}$. We say that the directed edge $\overrightarrow{(u, v)}$ is outgoing from $u$ and incoming to $v$. The number of directed edges coming into a vertex $v$ is called the indegree of $v$. Similarly, the number of edges going out from a vertex $u$ is called the outdegree of $u$. We draw a directed edge by an arc with an arrowhead. A directed graph is a graph whose all edges are directed. A simple graph has no loop and directed edges. Given a simple graph $G$ we can construct a directed graph $\overrightarrow{G}$ by assigning two opposite directions on every undirected edge. Therefore, if $(u, v)$ be an undirected edge in a simple graph $G$, we construct two directed edges $\overrightarrow{(u, v)}$ and $\overrightarrow{(v, u)}$ in $\overrightarrow{G}$.
	
	In this work, we utilize several simple graphs, such as path graphs, cycle graphs, star graphs, complete graphs, complete bipartite graphs, etc. which are depicted in figure \ref{simple_graphs}. If a graph has $n$ vertices we label them $v_0, v_1, \dots v_{n - 1}$. A path graph $P_n$ has edges $(v_0, v_1), (v_1, v_2), \dots (v_{(n - 2)}, v_{(n - 1)})$. Therefore, in $P_n$ degree of $v_0$ and $v_{(n - 1)}$ is $1$ and all other vertices have degree $2$. A cycle graph $C_n$ has edges $(u_0, u_1), (u_1, u_2), \dots (u_{n-1}, u_{0})$. Therefore, every vertex in $C_n$ has degree two. In a star graph $S_n$ $v_1, v_2, \dots v_{n - 1}$ are adjacent only to the central vertex $v_0$. Therefore, there are $(n - 1)$ vertices having degree $1$ and $v_0$ has degree $(n - 1)$. A complete graph $K_n$ is a graph in which every vertex is adjacent to all other vertices. Therefore, every vertex in $K_n$ has degree $(n - 1)$. The vertex set of a complete bipartite graph $K_{m,n}$ is partitioned into subsets $V_1$ and $V_2$, such that, there is no edge with both ends in the same partition. 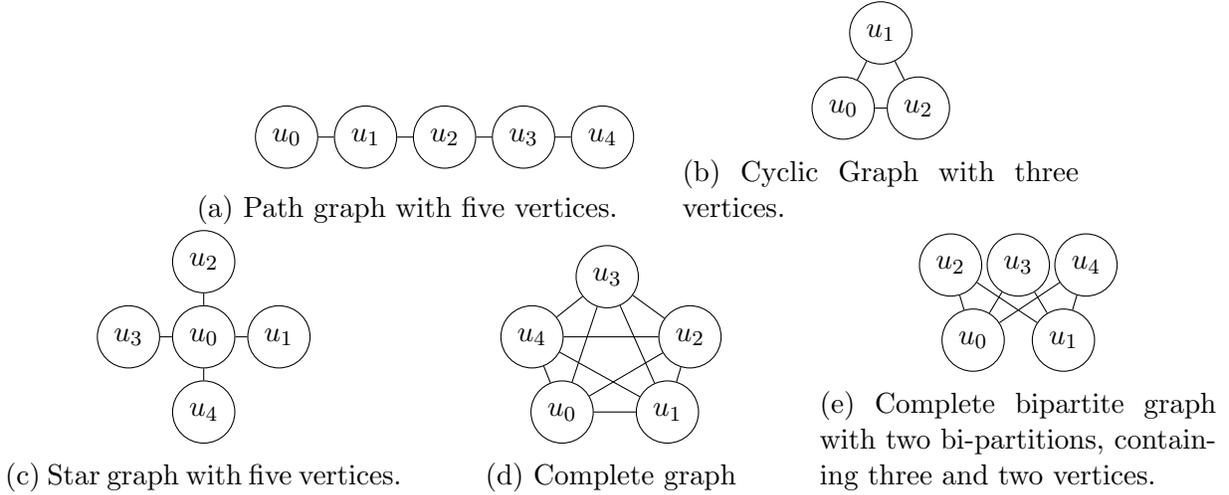
\begin{figure}
		\centering
		\begin{subfigure}{0.3\textwidth}
			\centering
			\begin{tikzpicture}[scale=1.4]
			\tikzstyle{every node}=[draw,shape=circle, font=\small,color=black];
			\path (0:0cm)  node (u_0) at(0,0) {$u_0$};
			\path (40:1cm)  node (u_1) at (0.7,0) {$u_1$};
			\path (100:1cm)  node (u_2) at (1.4,0) {$u_2$};
			\path (150:1cm)  node (u_3) at (2.1,0) {$u_3$};
			\path  (200:1cm) node (u_4) at (2.8,0) {$u_4$};
			\draw (u_0)--(u_1);
			\draw (u_1)--(u_2);
			\draw (u_2)--(u_3);
			\draw (u_3)--(u_4);
			\end{tikzpicture}
			\caption{Path graph $P_{5}$.}
			\label{fig:path_graph}
		\end{subfigure}
		\hspace{.2cm}
		\begin{subfigure}{0.3\textwidth}
			\centering
			\begin{tikzpicture}[scale=1.4]
			\tikzstyle{every node}=[draw,shape=circle,font=\small,color=black];
			\path (0:0cm)   node (u_0) at (0,0) {$u_0$};
			\path (36:1cm)  node (u_1) at (0.5,1) {$u_1$};
			\path (100:1cm)  node (u_2) at (1,0) {$u_2$};
			\draw (u_0)--(u_1);
			\draw (u_0)--(u_2);
			\draw (u_1)--(u_2);
			\end{tikzpicture}
			\caption{Cyclic Graph $C_{3}$.}
			\label{fig:cycle_graph}
		\end{subfigure}
		\hspace{.2cm}
		\begin{subfigure}{0.3\textwidth}
			\centering
			\begin{tikzpicture}[scale=1.2]
			\tikzstyle{every node}=[draw,shape=circle,font=\small,color=black];
			\path (0:0cm)  node (u_0)  at(0,0) {$u_0$};
			\path (40:1cm)  node (u_1)  at(1,0) {$u_1$};
			\path (100:1cm)  node (u_2)  at(0,1) {$u_2$};
			\path (150:1cm)  node (u_3)  at(-1,0) {$u_3$};
			\path  (200:1cm) node (u_4)  at(0,-1) {$u_4$};
			\draw (u_0)--(u_1);
			\draw (u_0)--(u_2);
			\draw (u_0)--(u_3);
			\draw (u_0)--(u_4);
			\end{tikzpicture}
			\caption{Star graph $S_{5}$.}
			\label{fig:star_graph}
		\end{subfigure}
		\\
		\begin{subfigure}{0.3\textwidth}
			\centering
			\begin{tikzpicture}[scale=1.4]
			\tikzstyle{every node}=[draw,shape=circle,font=\small,color=black];
			\path (0:0cm) node (u_0) at(-0.4,0) {$u_0$};
			\path (30:1cm) node (u_1) at(1,0) {$u_1$};
			\path (60:1cm) node (u_2) at(1.3,1)  {$u_2$};
			\path (90:1cm) node (u_3)  at(0.2,1.8) {$u_3$};
			\path (120:1cm) node (u_4)  at(-0.8,1) {$u_4$};
			\draw (u_0)--(u_1);
			\draw (u_0)--(u_2);
			\draw (u_0)--(u_3);
			\draw (u_0)--(u_4);
			\draw (u_1)--(u_2);
			\draw (u_1)--(u_3);
			\draw (u_1)--(u_4);
			\draw (u_2)--(u_3);
			\draw (u_3)--(u_4);
			\draw (u_2)--(u_4);
			\end{tikzpicture}
			\caption{Complete graph $K_{5}$.}
			\label{fig:complete_graph with five vertices}
		\end{subfigure}
		\hspace{.2cm}
		\begin{subfigure}{0.3\textwidth}
			\centering
			\begin{tikzpicture}[scale=1.5]
			\tikzstyle{every node}=[draw,shape=circle,font=\small];
			\path (0:0cm) node (u_0)  at(-0.2,0) [draw,shape=circle] {$u_0$};
			\path (30:1cm) node (u_1)  at(1,0) [draw,shape=circle,font=\small]  {$u_1$};
			\path (60:1cm) node (u_2)  at(-0.5,1) {$u_2$};
			\path (90:1cm) node (u_3)  at(0.4,1) {$u_3$};
			\path (120:1cm) node (u_4)  at(1.3,1) {$u_4$};
			\draw (u_0)--(u_2);
			\draw (u_0)--(u_3);
			\draw (u_0)--(u_4);
			\draw (u_1)--(u_2);
			\draw (u_1)--(u_3);
			\draw (u_1)--(u_4);
			\end{tikzpicture}
			\caption{Complete bipartite graph $K_{2,3}.$}
			\label{fig:complete_bi_graph}
		\end{subfigure}
		\hspace{.2cm}
		\begin{subfigure}{0.3\textwidth}
				\centering
				\begin{tikzpicture}[scale=1]
				\tikzstyle{every node}=[draw,font=\small,color=black];
				\node[draw, circle  ] (u_0) at (-0.4,0.5) {$u_0$};
				\node[draw, circle ] (u_1) at (-1.2,2) {$u_1$};
				\node[draw, circle ] (u_2) at (1.8,0.5) {$u_2$};
				\node[draw, circle ] (u_3) at (-1,-1) {$u_3$};
				\node[draw, circle ] (u_4) at (0.8,-1) {$u_4$};
				\node[draw, circle ] (u_5) at (3,-1) {$u_5$};
				\draw (u_0) -- (u_1);
				\draw (u_0) -- (u_2);
				\draw (u_0) -- (u_3);
				\draw (u_0) -- (u_4);
				\draw (u_2) -- (u_4);
				\draw (u_2) -- (u_5);
				\draw (u_3) -- (u_4);
				\draw (u_4) -- (u_5);
				\end{tikzpicture}
				\caption{An arbitrary graph with $6$ vertices $T_6$.}
				\label{fig:arbitrary_graph}
		\end{subfigure}
		\caption{A few graphs which we use an examples in this article.}
		\label{simple_graphs}
	\end{figure} 

	Corresponding to a graph $G$ with $n$ vertices, we choose an $n$-dimensional Hilbert space $\mathcal{H}^{(n)}$. A quantum state is a normalized vector $\ket{\psi}$ belonging to a Hilbert space, with one such vector associated with each vertex. The space $\mathcal{H}^{(n)}$ is spanned by the basis vectors $\ket{0}, \ket{1}, \dots \ket{(n - 1)}$. Recall that, a quantum state of dimension $2$ is called qubit. A quantum state is also represented by a density matrix. For a pure state $\ket{\psi}$, the density matrix is $\rho = \ket{\psi}\bra{\psi}$. In general, the density matrix of a quantum state is a positive semidefinite, Hermitian matrix with unit trace. 
	
	A quantum channel is used to transmit either classical or quantum information and can be represented mathematically by a set of Kraus operators \cite{nielsen2010quantum, choi1975completely, kraus2005operations}. Any set of Kraus operators $\{B_k\}$ representing a quantum channel must satisfy
	\begin{equation}\label{Kraus_operator_general}
	\sum_k B_k^\dagger B_k = I, 
	\end{equation}
	where $I$ is the identity operator. The action of the quantum channel on the state $\rho$ can be expressed as 
	\begin{equation}
	\rho'=\Lambda(\rho) = \sum_k B_k \rho B_k^\dagger,
	\end{equation}
	where $\rho'$ is the resulting state after the evolution. 
	
	The Weyl operators are the generalized Pauli operators in higher dimension \cite{bertlmann2008bloch}. There are multiple applications of the Weyl operators, particularly in quantum optics \cite{czerwinski2021quantum}, quantum computation and information theory \cite{man2006probabilistic,narnhofer2006entanglement}, as well as in quantum teleportation \cite{bennett1993teleporting,fonseca2019high}. The Weyl operators of order $n$ are defined by 
	\begin{equation}\label{Weyl_operator}
	U_{(u,v)}= \sum_{k=0}^{n-1} e^{(\frac{2\pi i}{n})ku} \ket{k}\bra{(k+v) \bmod n}, ~\text{where}~ 0 \leq v,u \leq (n-1).
	\end{equation}
	It can be proved that the operators $U_{(u,v)}$ are unitary for all $v$ and $u$, that is $U_{(u,v)}^\dagger U_{(u,v)} =  U_{(u,v)} U_{(u,v)}^\dagger = I_n$. When $u = v = 0$, we have $U_{(0, 0)} = I_n$. If $n = 2$ for $u = 1$ and $v = 0$, we have $U_{(1,0)} = \sigma_z$, which is the Pauli $Z$ matrix. Also, we obtain $\sigma_x$  and $i \sigma_y$ for $U_{(0,1)}$ and $U_{(1,1)}$, respectively.
	
	In this article, we study the non-Markovian behaviors of open quantum walk. One of the most common signatures of non-Markovian behaviour is characterized by a flow of information from the environment back to the open quantum system, which suggests the presence of memory effects \cite{de2017dynamics, breuer2016colloquium, laine2010measure, utagi2020temporal, utagi2020ping, paulson2021hierarchy, naikoo2019facets}. However, non-Markovian nature can be more complex in general \cite{shrikant2018non}. We use a number of measures of non-Markovianity to study the open system under our consideration. The fidelity between two quantum states \cite{liang2019quantum} represented by density matrices $\rho$ and $\sigma$ quantifies the distinguishability between them, and  is defined by $F(\rho , \sigma) = \left(\tr\sqrt{\sqrt{\rho}\sigma\sqrt{\rho}} \right)^2$, where $ 0 \leq F(\rho, \sigma) \leq 1 $. Also, $F(\rho , \sigma) = 1$ when $\rho = \sigma$. The dynamics of decoherence is one of the fundamental concerns in the study of open quantum systems. The quantum decoherence is the loss of quantum coherence in a state. Here, we use the $l_{1}$ norm of coherence to measure the quantum coherence of a state  \cite{chen2016quantifying, baumgratz2014quantifying, bhattacharya2018evolution} which is given by $C_{l_1}(\rho) = \sum_{i\neq j}|\rho_{ij}|$.

	\section{Construction of non-Markovian coin operators}\label{Section_for_coin_operators}
	
		Since we are interested in studying the impact of noise in DTQWs on various graphs, we introduce a coin operation that directly accommodates the relevant noisy channel.

		\subsection{Amplitude Damping Channel (non-Markovian)}
	
			The Amplitude Damping Channel (ADC) \cite{srikanth2008squeezed} describes the process by which a quantum system loses energy to its surroundings, that is energy dissipation of a quantum system to the environment. The non-Markovian ADC on an arbitrary $d$-dimensional system can be defined as follows \cite{dutta2023qudit}
			\begin{equation}
			C_0 = \ket{0}\bra{0} + \sqrt{1 - \lambda(t)} \sum_{u = 1}^{d - 1} \ket{u}\bra{u} \hspace{.5cm} \text{and} \hspace{.5cm} C_u = \sqrt{\lambda(t)}\ket{0}\bra{u}.
			\end{equation}
			Also, $\lambda(t) = 1 - e^{-gt}\left[ \frac{g}{l}\sinh\left(\frac{lt}{2}\right) + \cosh\left(\frac{lt}{2}\right) \right]^2$, where $l = \sqrt{g^2 - 2 \gamma g}$ and $g$ is the spectral width of the system-environment coupling and $\gamma$ is the spontaneous emission rate. Also, $\ket{0}$ and $\ket{u}: u = 1, 2, \dots (d - 1)$ are the $d$ dimensional state vectors in the computational basis of $\mathcal{H}^{(d)}$. Note that, $C_0$ and $C_u$ satisfy the conditions to be Kraus operators, as stated in equation (\ref{Kraus_operator_general}). 
			
			These operators are accommodated into a graph theoretic setting by applying them as the coin operators on the vertices of a graph. Let $u$ be a vertex with outdegree $d_u$. The outgoing edges from $u$ are $u_1, u_2, \dots u_{d_u}$. Also, we consider a loop $(u, u)$ at the vertex $u$. Therefore, we need $(d_u + 1)$ Kraus operators at $u$ each of dimension $n$, which are as follows
			\begin{equation}\label{ADC}
			\begin{split}
			C_{(u, u)} & = \ket{u}\bra{u} + \sqrt{1 - \lambda(t)} \sum_{\overrightarrow{(u, v)} \in E(\overrightarrow{G})} \ket{v}\bra{v} + \sum_{\overrightarrow{(u, w)} \notin 	E(\overrightarrow{G})} \ket{w}\bra{w},\\
			C_{(u, v)} & = \sqrt{\lambda(t)} \ket{u}\bra{v}.
			\end{split}
			\end{equation}
			Here, $\ket{u}, \ket{v}$, and $\ket{w}$ are state vectors in $\mathcal{H}^n$. We apply $C_{(u, u)}$ for the loop $(u, u)$ and $C_{(u, v)}$ for the edge $\overrightarrow{(u, v)}$. Here, both $C_{(u,u)}$ and $C_{(u,v)}$ are a matrices of order $n$. Also, $C_{(u,u)}$ is a diagonal matrix. The $(u,v)$-th position of $C_{(u,v)}$ is $\sqrt{\lambda}$ and the other entries are zero, for all graphs. As the parameter $\lambda(t)$ is a function of time, the coin operators are also the functions of time. They satisfy the completeness condition for being Kraus operators at each time-steps. For the detailed calculation the reader is referred to Appendix \ref{appendix}.

	\subsection{Non-Markovian Dephasing Channel}
	
		The Kraus operators representing the Non-Markovian Dephasing (NMD) noise for qudits \cite{dutta2023qudit} are as follows:
		\begin{equation}\label{Non-Markovian dephasing}
			C_{u,v} = \begin{cases} \sqrt{1 - \kappa(p)}~U_{(0,0)} & ~\text{when}~ u = 0, v = 0; \\ \sqrt{\frac{\kappa(p)}{d^2 - 1}}~U_{(u, v)} & ~\text{for}~ 0 \leq u, v \leq (d - 1) ~\text{and}~ 	(u, v) \neq (0, 0). \end{cases}
		\end{equation} 
	where
	\begin{equation}
	\kappa(p) = p \frac{1 + \eta (1 - 2p) \sin(\omega p)}{1 + \eta (1 - 2p)},
	\end{equation}
	and $0 \leq p \leq \frac{1}{2}$, as well as $\eta$ and $\omega$ are two constants characterizing channel strength and frequency. 
	
	In equation (\ref{Non-Markovian dephasing}) there are $(d^2 - 1)$ operators of the form $U_{u, v}$. We want to fit these operators with the outgoing edges from a vertex. Let $u$ be a vertex with outdegree $d_u$. Thus, we need $(d_u +1)$ coin operators associated to the outgoing edges from $u$. Hence, we modify the Kraus operators as follows:
	\begin{equation}\label{NMD}
	C_{(u, u)} = \sqrt{1 - \kappa(p)}~U_{(0,0)} \hspace{.5cm} \text{and} \hspace{.5cm} C_{(u, v)} = \sqrt{\frac{\kappa(p)}{d_u}} ~ U_{(u, v)}.
	\end{equation}
	Here, $C_{(u,u)}$ is the coin operator associated with the loop $(u,u)$ and $C{(u, v)}$ are the coin operators associated with the directed edges $(u, v)$. The operators $C_{(u,u)}$ form a diagonal matrix with diagonal elements $\sqrt{1-\kappa}$, for all graphs. These operators also satisfy the completeness criterion of Kraus operators. We discuss it in Appendix \ref{appendix}.
	
	\subsection{Non-Markovian Depolarization Channel}
	
		Depolarization noise also affects the quantum systems by causing the loss of information about their quantum states. For the qudits \cite{dutta2023qudit}, this noise is characterized by the following Kraus operators
	\begin{equation}
	\begin{split}
	C_{(u,u)} = & \sqrt{1 + \frac{(N^2 - 1)(1 - p)}{N^2}\Lambda_1}~ U_{(0, 0)},  ~\text{for}~ u=0; \\
	C_{(u,v)}= & {\sqrt \frac{p \Lambda_2}{N^2}}~ U_{(u, v)} ~\text{for}~ 0 \leq u, v \leq (N - 1), ~\text{and}~ (u, v) \neq (0, 0).
	\end{split}
	\end{equation}
	Here, $\Lambda_1 = -\alpha p$, $\Lambda_2 = \alpha(1 - p)$ and $(1-p)\Lambda_1 + p\Lambda_2 = 0$.
	
	In order to utilize the non-Markovian depolarization channel as a coin operator on graph vertices, we need to further generalize it. Hence, the Kraus operators for this channel are of the form, 
	\begin{equation}\label{Depolarization}
	C_{(u,u)} = \sqrt{1 + \frac{d_u(1-p)\Lambda_1}{d_u+1}}~ U_{(0,0)} \hspace{.5cm} \text{and} \hspace{.5cm} C_{(u,v)} = \sqrt{\frac{p\Lambda_2}{d_u+1}}~ U_{(u,v)},
	\end{equation}
	where $d_u$ is the outdegree of $u$. We apply $C_{(u, u)}$ for the loop $(u, u)$ and $C_{(u, v)}$ for the edge $\overrightarrow{(u, v)}$. In Appendix \ref{appendix}, we justify that these coin operators also satisfy the Kraus operator condition for every value of $\Lambda_1$ and $\Lambda_2$.

	\section{DTOQW on graphs}
	
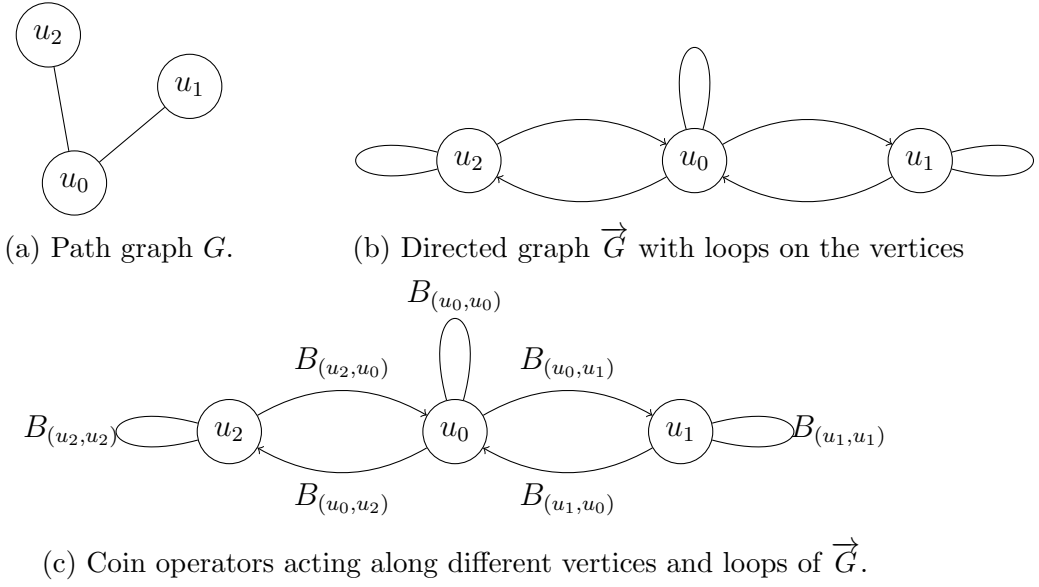
\begin{figure}
		\centering
		\begin{subfigure}{0.3\textwidth}
			\centering
			\begin{tikzpicture}[scale=2]
			\tikzstyle{every node}=[draw,shape=circle,color=black];
			\path (0:0cm)  node (u_0) {$u_0$};
			\path (40:1cm)  node (u_1) {$u_1$};
			\path (100:1cm)  node (u_2) {$u_2$};
			\draw (u_0)--(u_1);
			\draw (u_0)--(u_2);
			\end{tikzpicture}
			\caption{Path graph $G$.}
		\end{subfigure}
		\begin{subfigure}{0.5\textwidth} 
			\centering
			\begin{tikzpicture}
			\node[draw, circle] (u_0) at (0, 0) {$u_0$};
			\node[draw, circle] (u_1) at (3, 0) {$u_1$};
			\node[draw, circle] (u_2) at (-3,0) {$u_2$};
			\path (u_0) edge [loop above] ();
			\path (u_1) edge [loop right] ();
			\path (u_2) edge [loop left] (); 
			\path (u_0) edge [->, bend left] node[midway,above] {} (u_1);
			\path (u_1) edge [->, bend left] node[midway,above] {} (u_0);
			\path (u_0) edge [->, bend left] node[midway,above] {} (u_2);
			\path (u_2) edge [->, bend left] node[midway,above] {} (u_0);
			\end{tikzpicture}
			\caption{Directed graph $\overrightarrow{G}$ with loops on the vertices}
			\label{directed graph}
		\end{subfigure}
		\\
		\begin{subfigure}{0.8\textwidth}
			\centering
			\begin{tikzpicture}
			\node[draw, circle] (u_0) at (0, 0) {$u_0$};
			\node[draw, circle] (u_1) at (3, 0) {$u_1$};
			\node[draw, circle] (u_2) at (-3,0) {$u_2$};
			\path (u_0) edge [loop above] ();
			\path (u_1) edge [loop right] ();
			\path (u_2) edge [loop left] (); 
			\draw[->, bend left] (u_0) to node[midway, above] {$B_{(u_{0}, u_1)}$} (u_1);
			\draw[->, bend left] (u_1) to node[midway, below] {$B_{(u_{1}, u_0)}$} (u_0);
			\draw[->, bend left] (u_0) to node[midway, below] {$B_{(u_{0}, u_2)}$} (u_2);
			\draw[->, bend left] (u_2) to node[midway,above] {$B_{(u_{2},u_0)}$} (u_0);
			\node at (0, 1.8) {$B_{(u_{0}, u_0)}$};
			\node at (-5.1, 0) {$B_{(u_{2}, u_2)}$};
			\node at (5.1, 0) {$B_{(u_{1}, u_1)}$};
			\end{tikzpicture}
			\caption{Coin operators acting along different vertices and loops of $\overrightarrow{G}$.}
			\label{graph_with_operators}
		\end{subfigure}
		\caption{We consider a path graph $G$ with three vertices, in subfigure (a). We assign two opposite orientations on every edge of $G$ as well as add loops on the vertices to construct a directed graph, in subfigure (b). For every directed edge and loop, there is a coin operator, which is indicated in subfigure (c).}
		\label{example_graph}	
	\end{figure}

	To describe the dynamics of the walker on any simple graph $G$, we convert it to a directed graph $\overrightarrow{G}$ by assigning two opposite directions on every undirected edge. Also, we add loops to all the nodes in $\overrightarrow{G}$. Let $u$ be any vertex with degree $d_u$ in $G$. For simplicity, we can assume that there are directed edges $\overrightarrow{(u, u_1)}, \overrightarrow{(u, u_2)}, \dots \overrightarrow{(u, u_d)}$ and a loop $(u, u)$ incident to $u$ in the graph $\overrightarrow{G}$. We depict a graph $G$, and $\overrightarrow{G}$ in Figure \ref{example_graph}. 
	
	The discrete-time quantum walk is implemented by the coin and shift operators. We discussed the coin operators, in Section \ref{Section_for_coin_operators}. For every directed edge $\overrightarrow{(u, v)}$ we associate a coin operator $C_{(u,v)}$ and a shift operator $S_{(u,v)} = \ket{v}\bra{u}$. The transition operator for each edge $\overrightarrow{(u,u_{l})}$ for $l = 1, 2, \dots d_u$ is given by $B_{(u,u_{l})} = C_{(u,u_{l})} \otimes S_{(u,u_{l})}$. Also, for the loop $(u, u)$ we can define the transition operator as $B_{(u, u)} = C_{(u,u)} \otimes S_{(u,u)}$. Here, $\otimes$ denotes the tensor product between two operators. In Figure \ref{graph_with_operators}, we indicate the transition operators on a digraph $\overrightarrow{G}$. Now, we find a set of operators at each node $u$ which can lead a walker through the edges going out from it.
	As we assume $S_{(u,v)} = \ket{v}\bra{u}$. Therefore, $S_{(u,v)}^\dagger S_{(u,v)} = \ket{u}\braket{v|v}\bra{u} = \ket{u} \bra{u}. $
	
	For the transition operators associated with the outgoing edges from a vertex $u$, we find that
	\begin{equation}
	\begin{split}
	& B_{(u, u)}^\dagger B_{(u, u)} + \sum_{\overrightarrow{(u, v)} \in E(\overrightarrow{G})} B_{(u, v)}^\dagger B_{(u, v)} = C_{(u,u)}^\dagger C_{(u,u)} \otimes \ket{u}\bra{u} + \sum_{\overrightarrow{(u, v)} \in E(\overrightarrow{G})} C_{(u,v)}^\dagger C_{(u,v)} \otimes \ket{u}\bra{u}\\
	& = \left(C_{(u,u)}^\dagger C_{(u,u)} +  \sum_{\overrightarrow{(u, v)} \in E(\overrightarrow{G})} C_{(u,v)}^\dagger C_{(u,v)}\right) \otimes \ket{u}\bra{u} = I_n \otimes \ket{u} \bra{u}.\\
	\end{split}
	\end{equation}
	Now, considering all the vertices in $\overrightarrow{G}$ we observe that
	\begin{equation}\label{Kraus_operator_condition}
	\begin{split}
	& \sum_{u \in V(G)} \left[B_{(u, u)}^\dagger B_{(u, u)} + \sum_{\overrightarrow{(u, v)} \in E(\overrightarrow{G})} B_{(u, v)}^\dagger B_{(u, v)} \right] \\
	& = \sum_{u \in V(G)} I_n \otimes \ket{u} \bra{u} = I_n \otimes \sum_{u \in V(G)} \ket{u}\bra{u} = I_n \otimes I_n = I_{n^2}.\\
	\text{or}~ & \sum_{u \in V(G)} \sum_{\overrightarrow{(u, v)} \in E(\overrightarrow{G})\cup \{(u,u)\}} B_{(u,v)}^\dagger B_{(u,v)} = I_{n^2}.
	\end{split}
	\end{equation}
	Therefore, the set of all transition operators $B_{(u,v)}$ forms a set of Kraus operators on $\overrightarrow{G}$. The total number of $B_{(u,v)}$ operators is $(2m+n)$, where $m$ is the number of edges in $G$ and $n$ is the number of vertices in $G$. 
	
	Let the walker initiates walking from an arbitrarily chosen vertex $u$. The initial state of the walker is given by the density matrix.
	\begin{equation}
	\rho^{(0)} = \rho_{u}^{(0)} \otimes \ket{u} \bra{u} + \sum_{v \neq u \in V(G)} \rho_{v}^{(0)} \otimes \ket{v} \bra{v}.
	\end{equation} 
	Here we consider $\rho_{u}^{(0)} = \frac{J_n}{n}$ and $\rho_{v}^{(0)} = \frac{Z_n}{n}$, where $J_n$ is all-one matrix of order $n$ and $Z_n$ are all-zero matrices of order $n$. For simplicity, we assume that the walker starts moving from the node $u_{0}$ in all the considered graphs. Therefore, the initial state is 
	\begin{equation}\label{Initial_state}
		\rho^{(0)} = \frac{1}{n} J_n \otimes \ket{0} \bra{0}.
	\end{equation}
	The probability of getting the walker at vertex $u$ at time $t = 0$ is $\tr(\rho_{u}^{(0)})$. Now, the density matrix of the walker after $(k + 1)$-th step is given by
	\begin{equation}
	\begin{split}
	\rho^{(k+1)} & =  \Lambda(\rho^{(k)}) = \sum_{u \in V(G)} \sum_{\overrightarrow{(u, v)} \in E(\overrightarrow{G})\cup \{(u,u)\}} B_{(u,v)} \rho_{u}^{(k)} \otimes 	\ket{u} \bra{u} B_{(u,v)}^\dagger \\
	& = \sum_{u \in V(G)} \sum_{\overrightarrow{(u, v)} \in E(\overrightarrow{G})\cup \{(u,u)\}} C_{(u,v)} \rho_{u}^{(k)} C_{(u,v)}^\dagger \otimes \ket{v} \braket{u|u} 	\braket{u|u}\bra{v} \\
	& = \sum_{u \in V(G)} \left[\sum_{\overrightarrow{(u, v)} \in E(\overrightarrow{G})\cup \{(u,u)\}} C_{(u,v)} \rho_{u}^{(k)} C_{(u,v)}^\dagger \right] \otimes \ket{v} 	\bra{v}\\
	& = \sum_{u \in V(G)} \left[\sum_{\overrightarrow{(v, u)} \in E(\overrightarrow{G})\cup \{(v, v)\}} C_{(v,u)} \rho_{v}^{(k)} C_{(v, u)}^\dagger \right] \otimes \ket{u} 	\bra{u}. ~[\text{By changing the index $u$ and $v$.}]	
	\end{split}
	\end{equation}
	Comparing with equation (\ref{Initial_state}) we find that 
	\begin{equation}
	\rho_{u}^{(k + 1)} = \sum_{\overrightarrow{(v, u)} \in E(\overrightarrow{G})\cup \{(v, v)\}} C_{(v,u)} \rho_{v}^{(k)} C_{(v, u)}^\dagger.
	\end{equation}
	The probability of getting the walker at vertex $u$ at time-step $(k + 1)$ is $\tr(\rho_{u}^{(k + 1)})$. Below we discuss the quantum walk and its non-Markovian properties, characterized by coherence and fidelity, on the graphs depicted in Figure \ref{simple_graphs}.
	
	\subsection{DTOQW on path graphs} 
	
	Using equation (\ref{ADC}), we can derive the coin operators corresponding to the non-Markovian ADC channel for the path graph with $n$ vertices $P_n$. They are as follows:
	\begin{equation}
		\begin{split}
		& C_{(u,u)} = \begin{cases} 
				\ket{0}\bra{0} + \sqrt{1-\lambda(t)}\ket{1}\bra{1} + \sum_{w \neq 0}\ket{w} \bra{w} & ~\text{if}~ u = 0,\\
				\ket{n - 1}\bra{n - 1} + \sqrt{1-\lambda(t)}\ket{n - 2}\bra{n - 2} + \sum_{w \neq n - 1}\ket{w} \bra{w} & ~\text{if}~ u = (n - 1),\\
				\ket{u}\bra{u} + \sqrt{1-\lambda(t)}\left(\ket{u-1}\bra{u-1} + \ket{u+1}\bra{u+1} \right) \\ 
				\hspace{1cm} + \sum_{w \neq u, u \pm 1}\ket{w} \bra{w} & ~\text{if}~ u \neq 0 ~\text{and}~ u \neq (n - 1); \end{cases} \\
		&C_{(u,v)} = \sqrt{\lambda(t)}\ket{u}\bra{v}.
		\end{split}
	\end{equation}
	To define quantum walk, we also need the shift operators which are $S_{u,v} = \ket{v}\bra{u}$ for $u, v = 0, 1, \dots (n - 1)$. 
	
	For further calculation, we consider a path graph $P_5$ with five vertices as an example, which is depicted in Figure \ref{fig:path_graph}. The coin operator corresponding to the vertex $u_0$ using equation (\ref{ADC}) is
	\begin{equation}
	C_{(0,0)} = \ket{0}\bra{0} + \sqrt{1-\lambda(t)}\left (\ket{1}\bra{1}\right) + \ket{2}\bra{2} + 	\ket{3}\bra{3} + \ket{4}\bra{4}. 
	\end{equation}
	Also, vertex $u_1$ is adjacent to vertex $0$ only. Hence, the coin operator corresponding to vertex $u_1$ is
	\begin{equation}
	C_{(1,1)} = \ket{1}\bra{1} + \sqrt{1-\lambda(t)}\left (\ket{0}\bra{0}  + \ket{2}\bra{2}\right) + 	\ket{3}\bra{3} +  \ket{4}\bra{4}.
	\end{equation}
	Similarly, for the other loops the coin operators are
	\begin{equation}
	\begin{split}
	&C_{(2,2)} = \ket{2}\bra{2} + \sqrt{1-\lambda(t)}\left (\ket{1}\bra{1} + \ket{3}\bra{3} \right) + \ket{1}\bra{1} + \ket{4}\bra{4}, \\
	&C_{(3,3)} = \ket{3}\bra{3} + \sqrt{1-\lambda(t)}\left (\ket{2}\bra{2} + \ket{4}\bra{4} \right)  + 	\ket{1}\bra{1} + \ket{2}\bra{2}, \\
	&C_{(4,4)} = \ket{4}\bra{4} + \sqrt{1-\lambda(t)}\left (\ket{3}\bra{3}  \right) + \ket{0}\bra{0} + \ket{1}\bra{1} + \ket{2}\bra{2}.
	\end{split}
	\end{equation} 
	Also, the coin operators corresponding to the edges $(u,v)$ are $C_{(u,v)}  = \sqrt{\lambda(t)}\ket{u}\bra{v}$.
    
\begin{figure}
		\centering
		\includegraphics[scale = .5]{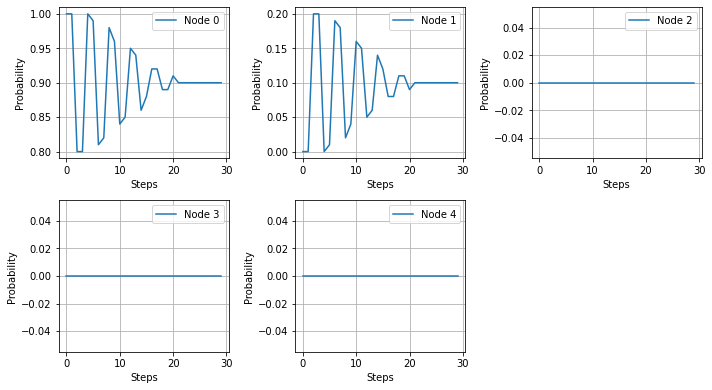}
		\caption{Probability distribution of the walker at different vertices of the path graph \ref{fig:path_graph} as a function of steps under the non-Markovian ADC taking $\gamma$ = $500$ and $g$ = $0.01$.}
		\label{Path_prob_ADC}
	\end{figure}

    We apply the DTOQW operators on the initial state discussed in equation (\ref{Initial_state}). 
	Figure \ref{Path_prob_ADC} indicates that the probability of getting the walker at different vertices of $P_5$ at different time steps. It suggests that the probability of getting the walker at vertex $1$ varies from $0.8$ to $1$. For the vertex $2$ this probability varies from $0$ to $0.2$. But, for the vertices $2$, $3$ and $4$ it remains $0$ in every steps. It suggests that the walker can not visit these nodes. Due to the limited connectivity of path graph the walker can move only to the neighbors of the initial vertex.
	
\begin{figure} 
		\begin{subfigure}{0.45\textwidth} 
			\centering
			\includegraphics[scale = .5]{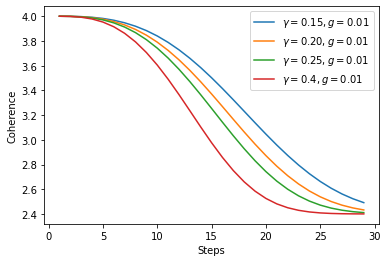}
			\caption{Coherence of the walker on a path graph $P_5$ as a function of time-steps for different values of $\gamma$ while $g = 0.01$.}
		\end{subfigure}
		\hfill
		\begin{subfigure}{0.45\textwidth}
			\centering
			\includegraphics[scale = .5]{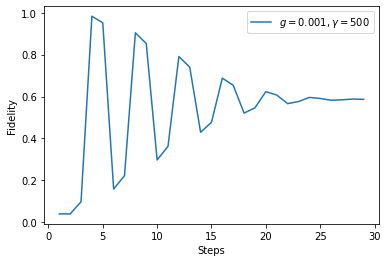}
			\caption{Fidelity $F(\rho^{(0)}, \rho^{(k)})$ of the walker on a path graph $P_5$ as a function of time-steps assuming $\gamma = 500$ and $g= 0.001$.}
		\end{subfigure}
		\caption{Coherence, and fidelity of a quantum walker on a path graph $P_5$ under non-Markovian ADC noise.}
		\label{Path_ADC}
	\end{figure}

	Figure \ref{Path_ADC} shows the coherence and the fidelity of a quantum walk on a path graph under non-Markovian ADC. Coherence decreases with increasing steps, indicating a steady loss of coherence over time. Lower $\gamma$ values result in a slower decline in coherence, whereas higher $\gamma$ values cause faster decoherence. Also, the fidelity fluctuates significantly before gradually stabilizing under the given parameter conditions.
	
\begin{figure}
		\centering
		\begin{subfigure}{0.32\textwidth} 
			\centering
			\includegraphics[scale = .5]{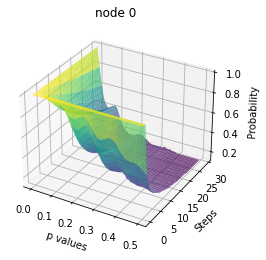}
		\end{subfigure}
		\begin{subfigure}{0.32\textwidth} 
			\centering
			\includegraphics[scale = .5]{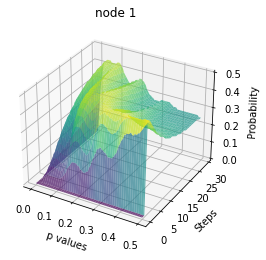}
		\end{subfigure}
		\begin{subfigure}{0.32\textwidth} 
			\centering
			\includegraphics[scale = .5]{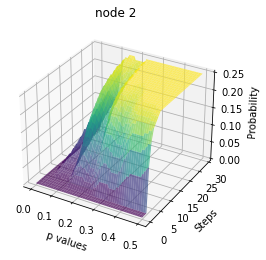}
		\end{subfigure}
		\begin{subfigure}{0.32\textwidth} 
			\centering
			\includegraphics[scale = .5]{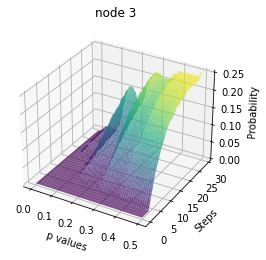}
		\end{subfigure}
		\begin{subfigure}{0.32\textwidth} 
			\centering
			\includegraphics[scale = .5]{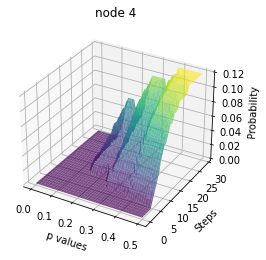}
		\end{subfigure}
		\caption{Probability distribution of the walker at different nodes of the path graph $P_5$ as a function of $p$ with time-steps under NMD noise assuming $\eta$ = $0.5$ and $\omega$ = $50$.}
		\label{Path_prob_NMD}
	\end{figure}

\begin{figure} 
		\begin{subfigure}{0.45\textwidth} 
			\centering
			\includegraphics[scale = .4]{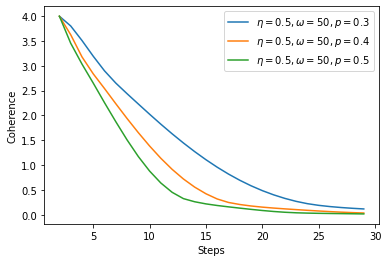}
			\caption{Coherence for path graph \ref{fig:path_graph} as a function of steps. Higher values of p result in a faster decline in coherence over steps.}
		\end{subfigure}
		\hfill
		\begin{subfigure}{0.45\textwidth}
			\centering
			\includegraphics[scale = .4]{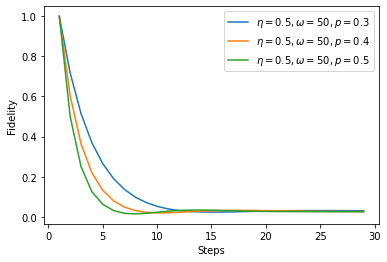}
			\caption{Fidelity for the path graph \ref{fig:path_graph} as a function of steps. The higher value of p results in a more rapid decline in fidelity over steps.}
		\end{subfigure}
		\caption{Coherence and fidelity of a quantum walker for the path graph $P_5$ under NMD noise.}
		\label{Path_NMD}
	\end{figure}

	Now, we consider the DTOQW using NMD channel on path graphs. In general using equation (\ref{NMD}), the coin operators are as follows:
	\begin{equation}\label{Coin_path_dephasing}
	\begin{split}
	&C_{(u,u)} = \sqrt{1-\kappa(p)}~ U_{(0, 0)}, \hspace{.5cm} C_{(0,1)} = \sqrt{\kappa(p)}~ U_{(0,1)}, \hspace{.5cm} C_{(n-1,n-2)} = \sqrt{\kappa(p)}~ U_{(n-1,n-2)},\\
	&C_{(u,v)} = \sqrt{\frac{\kappa(p)}{2}}~ U_{(u,v)}, ~\text{for}~ u \neq 0 ~\text{and}~ u \neq (n-1).
	\end{split}
	\end{equation}
	For a path graph $P_5$, we have $n = 5$. Then, the coin operators of DTOQW under NMD are defined as 
	\begin{equation}
		\begin{split}
			&C_{(u,u)} = \sqrt{1-\kappa(p)}~ U_{(0, 0)}, ~\text{for all}~ u = 0, 1, 2, 3, 4;\\
			&C_{(0,1)} = \sqrt{\kappa(p)} ~ U_{(0,1)}; \\
			&C_{(1,v)} = \sqrt{\frac{\kappa(p)}{2}}~ U_{(1,v)}, ~\text{for}~ v = 0 ~\text{and}~ 2;\\
			&C_{(2,v)} = \sqrt{\frac{\kappa(p)}{2}}~ U_{(2,v)}, ~\text{for}~ v = 1 ~\text{and}~ 3;\\
			&C_{(3,v)} = \sqrt{\frac{\kappa(p)}{2}}~ U_{(3,v)}, ~\text{for}~ v = 2 ~\text{and}~ 4;\\
			&C_{(4,3)} = \sqrt{\kappa(p)} ~ U_{(4,3)}.
		\end{split}
	\end{equation}
	The effects of these coin operators on the probability distribution of the walker, coherence and fidelity under NMD noise are shown in Figures \ref{Path_prob_NMD} and \ref{Path_NMD}, respectively. The probability of the walker decreases at node $0$ while it increases at the other nodes with steps and noise parameter $p$. Figure \ref{Path_NMD} suggests that the coherence and fidelity decrease with steps and channel parameters. 
	
	Next, we consider the DTOQW with the non-Markovian depolarization channel on path graph $P_5$. The coin operators are represented by
	\begin{equation}\label{path_coin}
		\begin{split}
			&C_{(u,u)} = \sqrt{1 + \frac{(1-p)\Lambda_1}{2}}~ U_{(0,0)}, ~\text{for}~ u = 0 ~\text{and}~ u = 4;  \\
			&C_{(u,u)} = \sqrt{1 + \frac{2(1-p)\Lambda_1}{3}}~ U_{(0,0)}, ~\text{for}~ u \neq 0 ~\text{and}~ u \neq 4;  \\
			&C_{(u,v)} = \sqrt{\frac{p\Lambda_2}{2}}~ U_{(u,v)}, ~\text{for}~ u= 0 ~\text{and}~ u = 4 ;\\ 
			&C_{(u,v)} = \sqrt{\frac{p\Lambda_2}{3}}~ U_{(u,v)}, ~\text{for}~ u \neq 0 ~\text{and}~ u \neq 4.
		\end{split} 
	\end{equation}

\begin{figure}
		\centering
		\includegraphics[scale = .5]{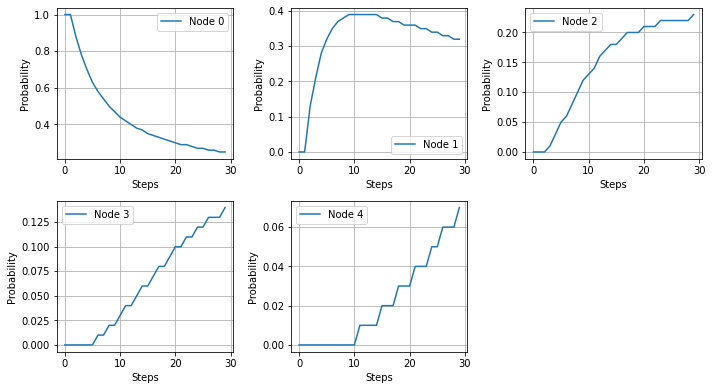}
		\caption{Probability distribution of the walker at different vertices for the path graph $P_5$ up to 30 steps under non-Markovian depolarization noise taking $p$ = $0.5$ and $\alpha$ = $1$.}
		\label{Path_prob_depolarization}
	\end{figure}

\begin{figure} 
		\begin{subfigure}{0.48\textwidth} 
			\centering
			\includegraphics[scale = .4]{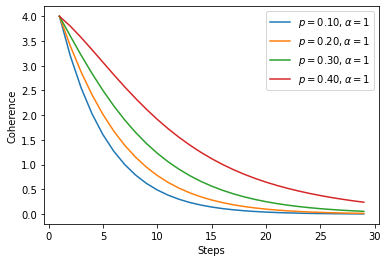}
			\caption{Coherence for the path graph \ref{fig:path_graph} as a function of time-steps and channel parameters $p$ and $\alpha$.}
		\end{subfigure}
		\hfill
		\begin{subfigure}{0.48\textwidth}
			\centering
			\includegraphics[scale = .4]{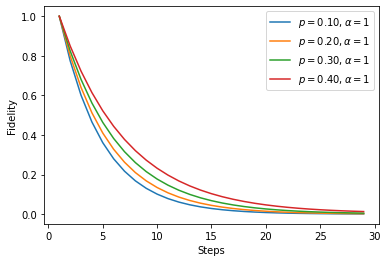}
			\caption{Here, also at a constant value of $\alpha$, the fidelity declines more rapidly with increasing p.}
		\end{subfigure}
		\caption{Coherence and fidelity of a quantum walker on a path graph $P_5$ under non-Markovian depolarization noise as a function of steps.}
		\label{Path_depolarization}
	\end{figure}

	Again, we apply these coin and shift operators on the initial density matrix (\ref{Initial_state}) of the walker and see the effect of these operators on the probability distribution, coherence and fidelity under non-Markovian depolarization noise, as depicted in Figures \ref{Path_prob_depolarization} and \ref{Path_depolarization}, respectively.  Here, the probability distribution for the path graph shows non-zero probability at all nodes because depolarization noise primarily affects the coherence and randomness, allowing the walker to maintain its ability to explore the graph and is shown in Figure \ref{Path_prob_depolarization}. The coherence and the fidelity decrease with the steps and $p$ while $\alpha$ is constant.  We observe that the coherence and the fidelity patterns for all the other considered graphs are qualitatively the same under the respective non-Markovian noises.
	
	\subsection{DTOQW on cycle graphs}
	
		Recall that a cycle graph $C_n$ with $n$ vertices has edges $(u_0, u_1), (u_1, u_2), \dots (u_{n-1}, u_{0})$. A cycle graph with three vertices is shown in Figure \ref{fig:cycle_graph}. We start our discussion of DTOQW on the cycle graph with the non-Markovian ADCthe non-Markovian ADC noise. Following equation (\ref{ADC}), the coin operators corresponding to the vertex $u$ are given by
		\begin{equation}
			\begin{split}
				&C_{(u,u)} = \ket{u}\bra{u} + \sqrt{1-\lambda(t)}\left (\ket{u-1}\bra{u-1} + \ket{u+2}\bra{u+2}\right) + \sum_{w \neq u}\ket{w} \bra{w}, \\
				&C_{(u,v)} = \sqrt{\lambda(t)}\ket{u}\bra{v}.
			\end{split}
		\end{equation} 
		
\begin{figure}
		\centering
		\includegraphics[scale = .5]{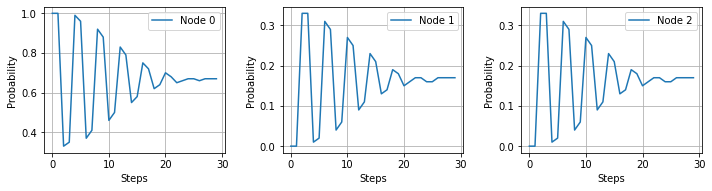}
		\caption{Probability distribution of the walker at different vertices of the cyclic graph $C_3$ as a function of steps under non-Markovian ADC. The channel parameters are the same as for path graph.}
		\label{Cyclic_prob_ADC}
	\end{figure}

		The probability distribution for the cycle graph $C_3$, which is depicted in Figure \ref{fig:cycle_graph}, are shown in Figures \ref{Cyclic_prob_ADC}. It can be observed from Figure \ref{Cyclic_prob_ADC} that due to connectivity and multiple paths, the walker can move to all the nodes, despite the effect of damping. We observe that the probability of the walker becomes stationary after around $25$ steps. Thereafter, there is no effect of the noise on the walker's probability with time-steps and an approximately stationary probability distribution is achieved. 
	
\begin{figure}
		\centering 
		\begin{subfigure}{0.32\textwidth} 
			\centering
			\includegraphics[scale = .5]{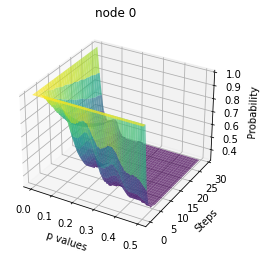}
		\end{subfigure}
		\hfil
		\begin{subfigure}{0.32\textwidth} 
			\centering
			\includegraphics[scale = .5]{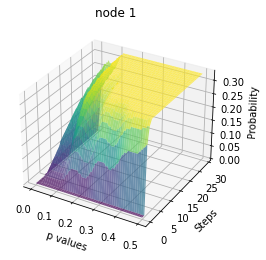}
		\end{subfigure}
		\hfil
		\begin{subfigure}{0.32\textwidth} 
			\centering
			\includegraphics[scale = .5]{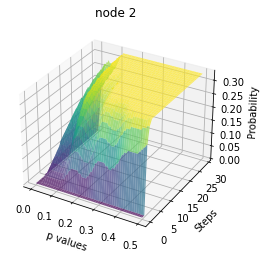}
		\end{subfigure}
		\caption{Probability distribution of the walker at different nodes for the cyclic graph $C_3$ as a function of p with steps under NMD noise. The channel parameters are the same as for path graph.} 
		\label{Cyclic_prob_NMD}
	\end{figure}

	The coin operators for non-Markovian dephasing channels can be defined as
	\begin{equation}
		C_{(u,u)} = \sqrt{1-\kappa(p)}~ U_{(0,0)} \hspace{.5cm} \text{and} \hspace{.5cm} C_{(u,v)} = \sqrt{\frac{\kappa(p)}{2}}~ U_{(u,v)}.
	\end{equation}
	 We apply these coin and shift operators on the DTOQW of the walker. The probability distribution of the walker is shown in figures \ref{Cyclic_prob_NMD}.  The probability of the walker at node $0$ decreases while at the other nodes of the cyclic graph increases with steps and $p$. 
	
\begin{figure}
		\centering
		\includegraphics[scale = .5]{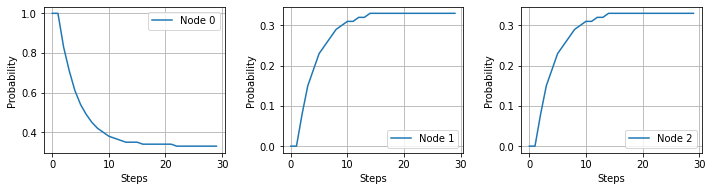}
		\caption{Probability distribution of the walker at different vertices for the cyclic graph \ref{fig:cycle_graph} with steps under non-Markovian depolarization noise. The channel parameters are the same as for path graph.}
		\label{Cyclic_prob_depolarization}
	\end{figure}

	The coin operators for the non-Markovian depolarization channel for the cyclic graph are
	\begin{equation}
		C_{u,u} = \sqrt{1 + \frac{2(1-p)\Lambda_1}{3}}~U_{0,0} \hspace{.5cm} \text{and} \hspace{.5cm} C_{u,v} = \sqrt{\frac{p\Lambda_2}{3}}~U_{(u,v)}.
	\end{equation}
	We apply these coin and shift operators on the DTOQW of the walker. In Figure \ref{Cyclic_prob_depolarization}, we observe the probability of the walker on all vertices due to the connectivity on the Cycle graph. 
	
	\subsection{DTOQW on star graphs}
	
	Recall that a star graph with $n$ vertices $S_n$ has a central vertex $u_0$. Each of the vertices $u_1$, $u_2$, $\dots$ $u_{(n - 1)}$ are adjacent to $u_0$ only. A star graph with $5$ vertices $S_5$ is shown in Figure \ref{fig:star_graph}. In a star graph, the walker can jump only between adjacent vertices. 
	
\begin{figure}
		\centering
		\includegraphics[scale = .5]{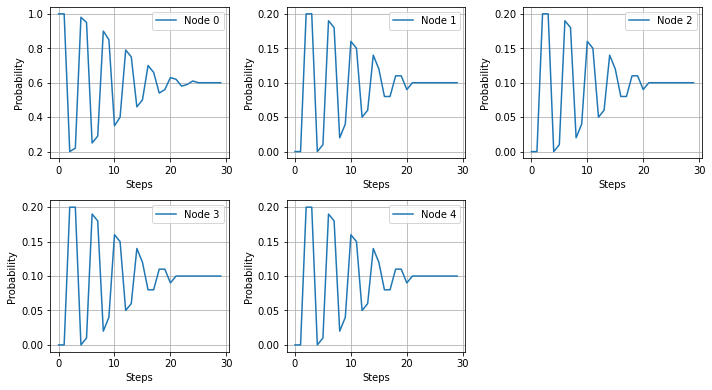}
		\caption{Probability distribution of the walker for the star graph \ref{fig:star_graph} at various vertices as a function of steps under non-Markovian ADC. The channel parameters are the same as for path graph.}
		\label{Star_prob_ADC}	
	\end{figure}

	The coin operators for the star graph $S_n$ under the non-Markovian ADC noise are 
	\begin{equation}
	\begin{split}
	& C_{(0,0)} = \ket{0}\bra{0} + \sqrt{1-\lambda(t)}\left (\sum_{r \neq 0}^{n-1} \ket{r}\bra{r} \right), \\
	& C_{(u,u)} = \ket{u} \bra{u} + \sqrt{1-\lambda(t)}\ket{0}\bra{0} + \sum_{w \neq u}\ket{w} \bra{w},\\
	& C_{(u,v)} = \sqrt{\lambda(t)}\ket{u}\bra{v}.
	\end{split}
	\end{equation}
	 Here, non-Markovian ADC noise shows different behaviour as for the star graph, the central vertex is connected to all other vertices thereby enhancing the connectivity. The probability at the node $u_0$ starts decreasing and at the other nodes starts increasing with steps due to this type of structure and the probability distribution becomes stationary after some steps, as can be seen from Figure \ref{Star_prob_ADC}.
	
	The coin operators for the star graph under the NMD channel are given by 
	\begin{equation}\label{star_coin_dephasing}
	\begin{split}
	&C_{(u,u)} = \sqrt{1-\kappa(p)}~U_{(0,0)};\\
	&C_{(u,v)} = \sqrt{\frac{\kappa(p)}{n-1}}~ U_{(u,v)},~\text{for}~ u=0; \\
	&C_{(u,v)} = \sqrt{\kappa(p)}~U_{(u,v)}, ~\text{for}~ u \neq 0.
	\end{split}
	\end{equation}	
	
\begin{figure}
		\centering 
		\begin{subfigure}{0.32\textwidth} 
			\centering
			\includegraphics[scale = .5]{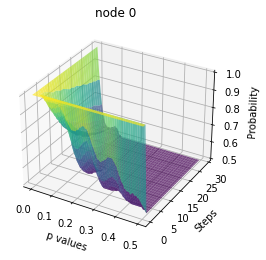}
		\end{subfigure}
		\begin{subfigure}{0.32\textwidth} 
			\centering
			\includegraphics[scale = .5]{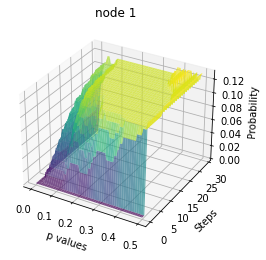}
		\end{subfigure}
		\begin{subfigure}{0.32\textwidth} 
			\centering
			\includegraphics[scale = .5]{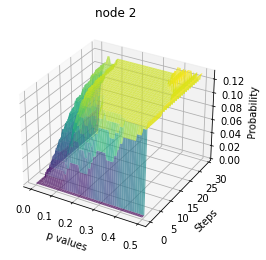}
		\end{subfigure}
		\begin{subfigure}{0.32\textwidth} 
			\centering
			\includegraphics[scale = .5]{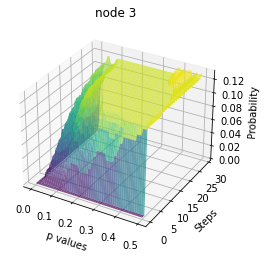}
		\end{subfigure}
		\begin{subfigure}{0.32\textwidth} 
			\centering
			\includegraphics[scale = .5]{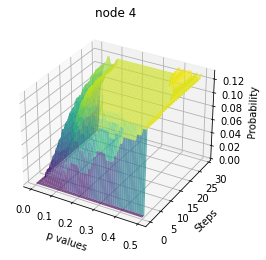}
		\end{subfigure}
		\caption{Probability distribution with time steps of the walker at different vertices of the star graph $S_5$, which is depicted in Figure \ref{fig:star_graph}, under NMD noise. The channel parameters are the same as for path graph.}
		\label{Star_prob_NMD}
	\end{figure}

     The probability distribution for getting the walker at different vertices of $S_5$ under NMD noise is shown in Figures \ref{Star_prob_NMD}. The probability of the walker at the node $0$ on the star graph decreases while increasing at the other nodes with steps and $p$.
	
\begin{figure}
		\centering
		\includegraphics[scale = .5]{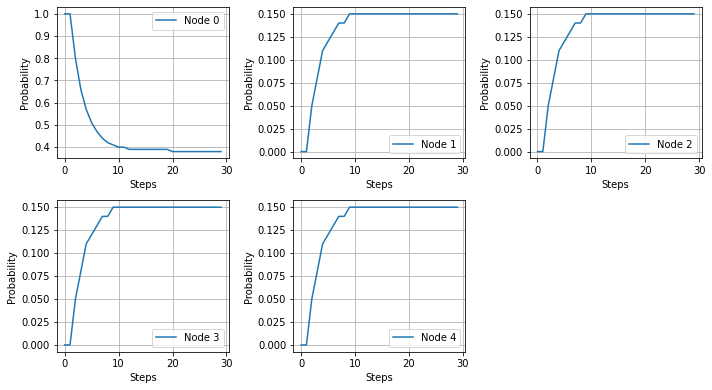}
		\caption{Probability distribution of the walker at different vertices of the star graph $S_5$ with steps under non-Markovian depolarization noise. The channel parameters are the same as for path graph.}
		\label{Star_prob_depolarization}
	\end{figure}

	The coin operators for the star graph under a non-Markovian depolarization channel can be described as
	\begin{equation}
	\begin{split}
	&C_{0,0} = \sqrt{1 + \frac{(n-1)(1-p)\Lambda_1}{n}}~ U_{(0,0)}; \\
	&C_{(u,u)} = \sqrt{1 + \frac{(1-p)\Lambda_1}{2}}~ U_{(0,0)}, ~\text{for}~ u \neq 0;  \\ 
	&C_{(0,u)} = \sqrt{\frac{p\Lambda_2}{n}}~ U_{(0,u)};\\
	&C_{(u,0)} = \sqrt{\frac{p\Lambda_2}{2}}~ U_{(u,0)}, ~\text{for}~ u \neq 0.
	\end{split}  
	\end{equation}
	The observed probability distribution on the star graph under non-Markovian depolarization noise is shown in Figure \ref{Star_prob_depolarization}. We can observe in Figure  \ref{Star_prob_depolarization} that depolarization noise causes the probability at vertex $0$ to decrease with respect to the time steps. Also, the probability at peripheral vertices (vertices $1$ to $4$) increases. Over the time, the probabilities stabilize, reflecting a stationary distribution across all nodes due to the randomizing effect of non-Markovian depolarization noise. Due to the structure of the star graph, the probability for the other nodes in the graph increases up to $10$ steps and then becomes stationary.
	
	\subsection{DTOQW on complete graphs}
	
	Recall that a complete graph consists of all possible edges between the vertices. A complete graph $K_5$ with five vertices is shown in Figure \ref{fig:complete_graph with five vertices}. 
	
	The non-Markovian ADC coin operators for $K_n$ are
	\begin{equation}
	C_{(u,u)} = \ket{u}\bra{u} + \sqrt{1-\lambda(t)}\sum_{r \neq u}^{n-1} \ket{r}\bra{r}  \hspace{.5cm} \text{and} \hspace{.5cm} C_{(u,v)} = \sqrt{\lambda(t)}\ket{u}\bra{v}.
	\end{equation} 			  
	
\begin{figure}
		\centering
		\includegraphics[scale = .5]{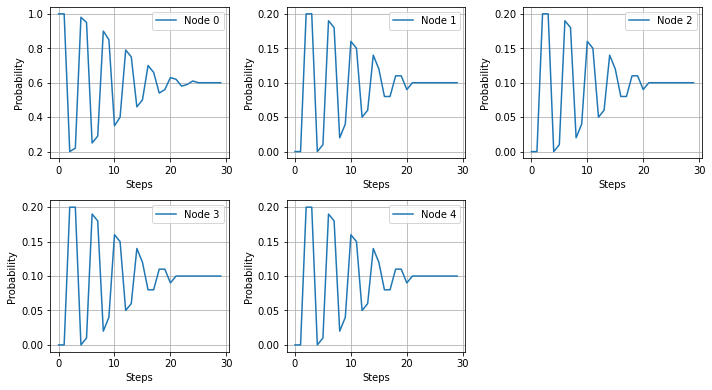}
		\caption{Under non-Markovian ADC, the walker's probability distribution at various vertices for the complete graph $K_5$ as a function of steps. The channel parameters are the same as for path graph.}
		\label{Complete_prob_ADC}
	\end{figure}

	Probability distribution plots are shown in Figures \ref{Complete_prob_ADC} for $K_5$ under non-Markovian ADC. In Figure \ref{Complete_prob_ADC}, initial oscillations in the probabilities are due to high connectivity, gradually dampening over time and becoming stationary after some steps.
	
\begin{figure}
		\centering
		\begin{subfigure}{0.32\textwidth} 
			\centering
			\includegraphics[scale = .5]{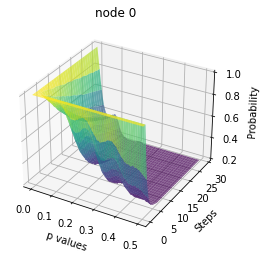}
		\end{subfigure}
		\begin{subfigure}{0.32\textwidth} 
			\centering
			\includegraphics[scale = .5]{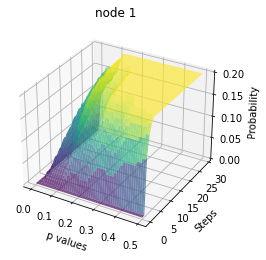}
		\end{subfigure}
		\begin{subfigure}{0.32\textwidth} 
			\centering
			\includegraphics[scale = .5]{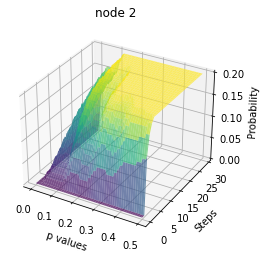}
		\end{subfigure}
		\begin{subfigure}{0.32\textwidth} 
			\centering
			\includegraphics[scale = .5]{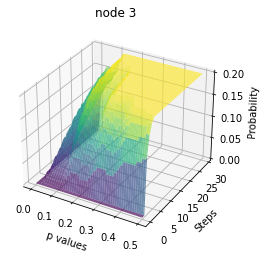}
		\end{subfigure}
		\begin{subfigure}{0.32\textwidth} 
			\centering
			\includegraphics[scale = .5]{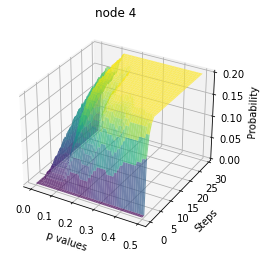}
		\end{subfigure}
		\caption{Probability distribution of the walker at different vertices of the complete graph $K_5$ under NMD noise. The channel parameters are the same as for path graph.}
		\label{Complete_prob_NMD}
	\end{figure}

	The non-Markovian dephasing coin operators for the complete graph can be described as follows:
	\begin{equation}
	C_{(u,u)} = \sqrt{1-\kappa(p)}~U_{(0,0)} \hspace{.5cm} \text{and} \hspace{.5cm} C_{(u,v)} = \sqrt{\frac{\kappa(p)}{n-1}}~U_{(u,v)}.
	\end{equation}
	The plots for the probability distribution of the walker on the graph under NMD are shown in Figures \ref{Complete_prob_NMD}. The probability plots show that the probability decreases over steps with oscillations due to memory effects at vertex $0$ while the probability of a quantum walker increases over steps at vertices $1$, $2$, $3$ and $4$. 
	
\begin{figure}
		\centering
		\includegraphics[scale = .5]{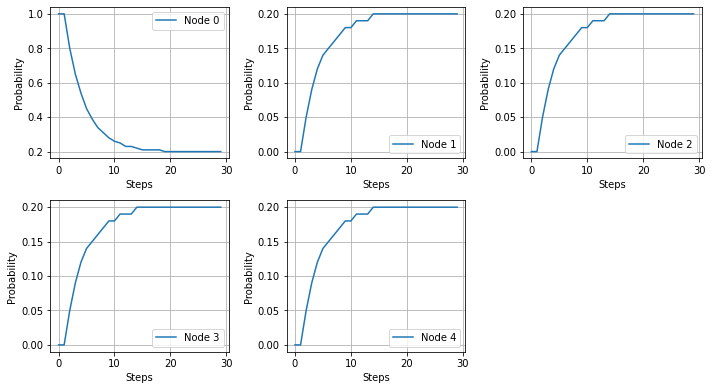}
		\caption{Probability distribution of the walker on the complete graph $K_5$ which influences the position of the walker over steps under non-Markovian depolarization noise. The channel parameters are the same as for path graph.}
		\label{Complete_prob_depolarization}
	\end{figure}

	The coin operators for the complete graph under non-Markovian depolarization channel can be described as,
	\begin{equation}
	C_{(u,u)} = \sqrt{1 + \frac{(n-1)(1-p)\Lambda_1}{n}}~ U_{(0,0)} \hspace{.5cm}\text{and} \hspace{.5cm} C_{(u,v)} = \sqrt{\frac{p\Lambda_2}{n}}~U_{(u,v)}.
	\end{equation}
	Here, the coin operators for loops $C_{(u,u)}$ form a diagonal matrix with equal entries $\sqrt{1 + \frac{(n-1)(1-p)\Lambda_1}{n}}$. 
	The probability distribution of the walker for the complete graph $K_5$ under non-Markovian depolarization noise is depicted in Figures \ref{Complete_prob_depolarization}. The probability plots show that the probability of the walker at the node $0$ is decreasing while increasing with equal distribution at the other nodes with the steps due to the structure of the complete graph.

	\subsection{DTOQW on complete bipartite graphs}
	
		Recall that, the vertex set of a complete bipartite graph is partitioned into two subsets $V_1$ and $V_2$. Also, every edge joins a vertex of $V_1$ and another vertex in $V_2$. An example of a complete bipartite graph is depicted in Figure \ref{fig:complete_bi_graph}, where $V_1 = \{u_{0}, u_{1}\}$ and  $V_2 = \{u_{2}, u_{2}, u_{3}, u_{4}\}$.
	
		The general form of coin operators for the complete bipartite graph under non-Markovian ADC is
		\begin{equation}
			\begin{split}
			&C_{(u,u)} = \sum_{u=0}^m \ket{u}\bra{u} + \sqrt{1-\lambda(t)}\sum_{r = m + 1}^{n-1} \ket{r}\bra{r}, ~ {u \in V_1 ~\text{and} ~ r \in V_2};\\
			&C_{(u,u)} = \sqrt{1-\lambda(t)} \sum_{u=0}^m \ket{u}\bra{u} +  \sum_{r = m + 1}^{n-1} 	\ket{r}\bra{r}, ~ {u \in V_1 ~ \text{and} ~ r \in V_2}; \\
			&C_{(u,v)} = \sqrt{\lambda(t)}\ket{u}\bra{v}.
			\end{split}
		\end{equation} 
	
\begin{figure}
		\centering
		\includegraphics[scale = .5]{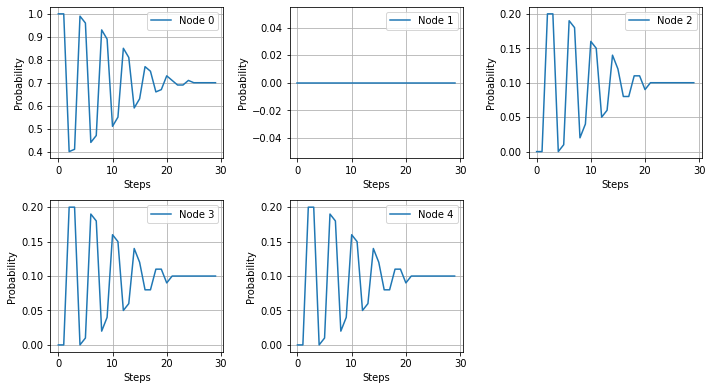}
		\caption{Walker's probability distribution at various vertices for the complete bipartite graph as a function of steps under non-Markovian ADC. The channel parameters are the same as for path graph.}
		\label{Complete_bi_prob_ADC}
	\end{figure}

		The probability distribution for the complete bipartite graph $K_{2,3}$, depicted in \ref{fig:complete_bi_graph}, under non-Markovian ADC is in Figures \ref{Complete_bi_prob_ADC}. The probability distribution of the walker on the complete bipartite graph decreases at vertex $0$ with steps and then becomes stationary. Vertices $u_2$, $u_3$, and $u_4$ show increasing probabilities with oscillatory patterns initially, thereafter it becomes stationary with steps. The probability at vertex $1$ remains negligible due to its initial disconnected state.
	
\begin{figure}
		\centering
		\begin{subfigure}{0.32\textwidth} 
			\centering
			\includegraphics[scale = .5]{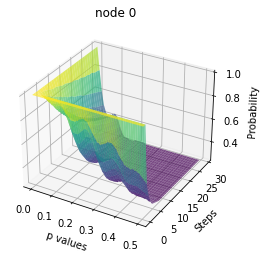}
		\end{subfigure}
		\begin{subfigure}{0.32\textwidth} 
			\centering
			\includegraphics[scale = .5]{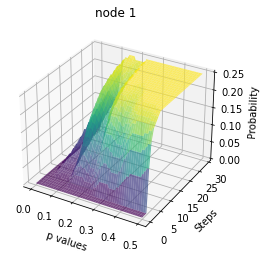}
		\end{subfigure}
		\begin{subfigure}{0.32\textwidth} 
			\centering
			\includegraphics[scale = .5]{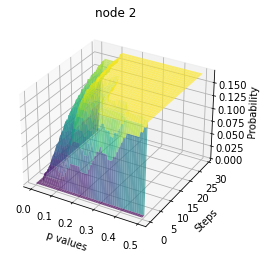}
		\end{subfigure}
		\begin{subfigure}{0.32\textwidth} 
			\centering
			\includegraphics[scale = .5]{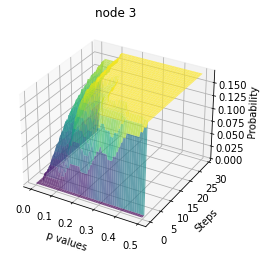}
		\end{subfigure}
		\begin{subfigure}{0.32\textwidth} 
			\centering
			\includegraphics[scale = .5]{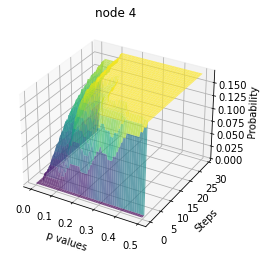}
		\end{subfigure}
		\caption{Probability distribution of the walker at all nodes of the complete bipartite graph $K_{2.3}$ under NMD noise. The channel parameters are the same as for path graph.}
		\label{complete_bi_prob_NMD}
	\end{figure}

	The coin operators for the general complete bipartite graph under a non-Markovian dephasing channel are defined by
	\begin{equation}
	\begin{split}
	&C_{(u,u)} = \sqrt{1-\kappa(p)}~U_{(0,0)};\\
	&C_{(u,v)} = \sqrt{\frac{\kappa(p)}{n-m}}~U_{(u,v)}, ~ {u \in V_1 ~\text{and} ~ v \in V_2}; \\
	&C_{(u,v)} = \sqrt{\frac{\kappa(p)}{m}}~U_{(u,v)}, ~ {u \in V_2 ~\text{and} ~ v \in V_1}.
	\end{split}
	\end{equation}
	The probability distribution of the walker on the complete bipartite graph under NMD noise is shown in the Figures \ref{complete_bi_prob_NMD}. The probability of the walker decreases at vertex $u_0$ while increases at the other nodes with steps and becomes stationary after $20$ steps.
	
\begin{figure}
		\centering
		\includegraphics[scale = .5]{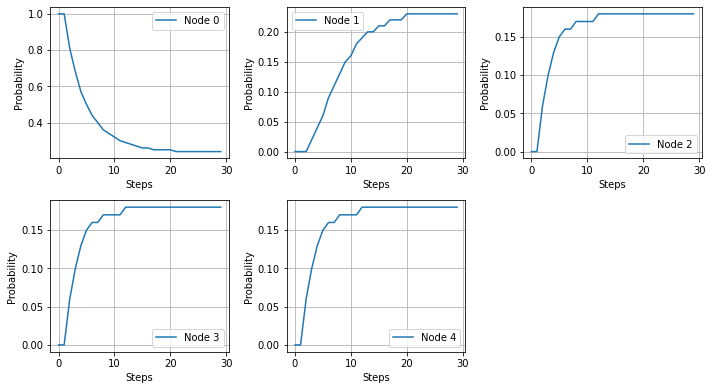}
		\caption{Probability distribution of the walker on the graph $K_{2,3}$ under the non-Markovian depolarization noise. The channel parameters are the same as for path graph.}
		\label{complete_bi_prob_depolarizaton}
	\end{figure}

	The coin operators for the complete bipartite graph under non-Markovian depolarization noise, can be characterized as,
	\begin{equation}
	\begin{split}
	&C_{(u,u)} = \sqrt{1 + \frac{(n-m)(1-p)\Lambda_1}{n-m+1}}~ U_{(0,0)},~ \text{if}~ {u \in V_1 };\\
	&C_{(u,u)} = \sqrt{1 + \frac{m(1-p)\Lambda_1}{m+1}}~ U_{(0,0)}, ~ \text{if} ~ {u \in V_2 };\\
	&C_{(u,v)} = \sqrt{\frac{p\Lambda_2}{n-m+1}}~U_{(u,v)}, ~ \text{if}~ {u \in V_1 ~\text{and} ~ v \in V_2};\\ 
	&C_{(u,v)} = \sqrt{\frac{p\Lambda_2}{m+1}}~U_{(u,v)},~ \text{if}~ {u \in V_2 ~\text{and} ~ v \in V_1}.
	\end{split}
	\end{equation}
	The probability distribution of the walker on the complete bipartite graph under non-Markovian depolarization noise is depicted in Figure \ref{complete_bi_prob_depolarizaton}. The probability of the walker decreases at vertex $0$ while increases at the other nodes with steps and becomes stationary after some steps. 

	\subsection{DTOQW on an arbitrary graph} 
		
		In this subsection, we consider a graph with $6$ vertices which does not belong to any of the above classes. We denote this graph with $T_6$ and depict it in Figure \ref{fig:arbitrary_graph}.
		
		We start our discussion of DTOQW under the influence of non-Markovian ADC noise. Note that, vertex $0$ is adjacent to the vertices $1, 2, 3$, and $4$. Therefore, following equation (\ref{ADC}), the coin operator corresponding to vertex $0$ is
		\begin{equation}
			C_{(0,0)} = \ket{0}\bra{0} + \sqrt{1-\lambda(t)}\left (\ket{1}\bra{1} + \ket{2}\bra{2} + 	\ket{3}\bra{3} + \ket{4}\bra{4} \right) + \ket{5}\bra{5}.
		\end{equation}
		Also, vertex $1$ is adjacent to vertex $0$ only. Hence, the coin operator corresponding to vertex $1$ is
		\begin{equation}
			C_{(1,1)} = \ket{1}\bra{1} + \sqrt{1-\lambda(t)}\left (\ket{0}\bra{0} \right) + \ket{2}\bra{2} + 	\ket{3}\bra{3} +  \ket{4}\bra{4} + \ket{5}\bra{5}.
		\end{equation}
		Similarly, we can construct the other coin operators, which are
		\begin{equation}
			\begin{split}
				&C_{(2,2)} = \ket{2}\bra{2} + \sqrt{1-\lambda(t)}\left (\ket{0}\bra{0} + \ket{4}\bra{4} + 	\ket{5}\bra{5} \right) + \ket{1}\bra{1} + \ket{3}\bra{3}, \\
				&C_{(3,3)} = \ket{3}\bra{3} + \sqrt{1-\lambda(t)}\left (\ket{0}\bra{0} + \ket{4}\bra{4} \right)  + 	\ket{1}\bra{1} + \ket{2}\bra{2} + \ket{5}\bra{5}, \\
				&C_{(4,4)} = \ket{4}\bra{4} + \sqrt{1-\lambda(t)}\left (\ket{0}\bra{0} + \ket{2}\bra{2} + 	\ket{3}\bra{3} \right) + \ket{1}\bra{1} + \ket{5}\bra{5}, \\
				&C_{(5,5)} = \ket{5}\bra{5} + \sqrt{1-\lambda(t)}\left (\ket{2}\bra{2} + \ket{4}\bra{4} \right) + 	\ket{0}\bra{0} + \ket{1}\bra{1} + \ket{3}\bra{3}.
			\end{split}
		\end{equation} 
		Corresponding to the edges $(u,v)$ there are the coin operators $C_{(u,v)} = \sqrt{\lambda(t)}\ket{u}\bra{v}$. The shift operators are $ S_{(u,v)} = \ket{v}\bra{u} $.  
		
\begin{figure}
		\centering
		\includegraphics[scale = .5]{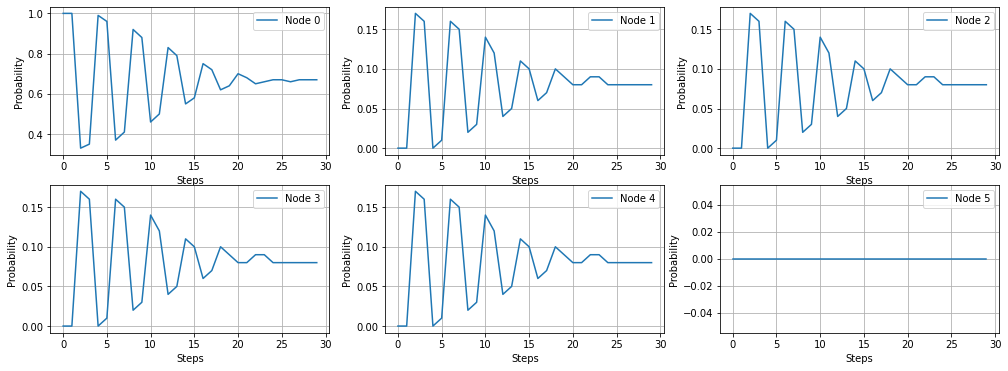}
		\caption{Under non-Markovian ADC, the walker's probability distribution at various vertices for $T_{6}$ as a function of steps. The channel parameters are the same as for path graph.}
		\label{Irregular_prob_ADC}
	\end{figure}

		The walker is assumed to move from the node $u_{0}$ on the $T_{6}$ graph. The initial state of the walker is given by the equation (\ref{Initial_state}). The probability distribution for the graph $T_{6}$ under non-Markovian ADC is depicted in Figure \ref{Irregular_prob_ADC}. We observe that due to the connectivity and ADC noise, the probability of the walker changes at neighbouring vertices with time-steps except vertex $5$.
	
	 The coin operators for the graph $T_{6}$ under a non-Markovian dephasing channel are defined by
	\begin{equation}
	\begin{split}
	&C_{(u,u)} = \sqrt{1-\kappa(p)}~ U_{(0,0)}; \\
	&C_{(0,v)} = \sqrt{\frac{\kappa(p)}{4}} ~ U_{(0,v)}, ~\text{for}~ 1 \leq  v \leq (n - 2); \\
	&C_{(1,0)} = \sqrt{\kappa(p)}~ U_{(1,0)}; \\
	&C_{(2,v)} = \sqrt{\frac{\kappa(p)}{3}}~ U_{(2,v)}, ~\text{for}~ v =0, 4 ~\text{and} ~ 5; \\
	&C_{(3,v)} = \sqrt{\frac{\kappa(p)}{2}}~ U_{(3,v)}, ~\text{for}~ v =0 ~\text{and} ~ 4;\\
	&C_{(4,v)} = \sqrt{\frac{\kappa(p)}{4}}~ U_{(4,v)}, ~\text{for}~ v =0, 2, 3 ~\text{and} ~ 5; \\
	&C_{(5,v)} = \sqrt{\frac{\kappa(p)}{2}}~ U_{(5,v)}, ~\text{for}~ v =2 ~\text{and}~ 4.
	\end{split}
	\end{equation} 

\begin{figure}
	\centering
	\begin{subfigure}{0.32\textwidth} 
		\centering
		\includegraphics[scale = .5]{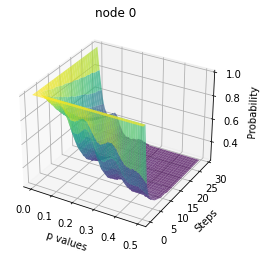}
	\end{subfigure}
	\begin{subfigure}{0.32\textwidth} 
		\centering
		\includegraphics[scale = .5]{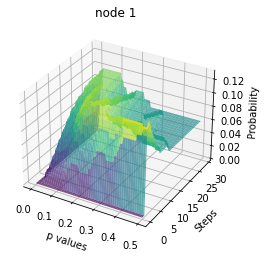}
	\end{subfigure}
	\begin{subfigure}{0.32\textwidth} 
		\centering
		\includegraphics[scale = .5]{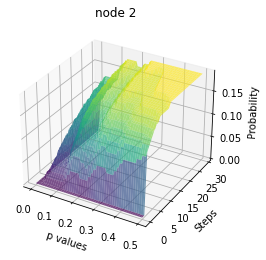}
	\end{subfigure}
	\begin{subfigure}{0.32\textwidth} 
		\centering
		\includegraphics[scale = .5]{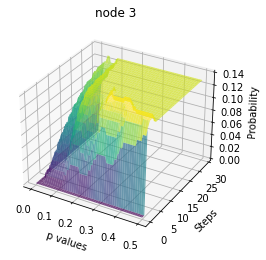}
	\end{subfigure}
	\begin{subfigure}{0.32\textwidth} 
		\centering
		\includegraphics[scale = .5]{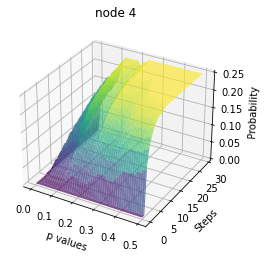}
	\end{subfigure}
    \begin{subfigure}{0.32\textwidth} 
    	\centering
    	\includegraphics[scale = .5]{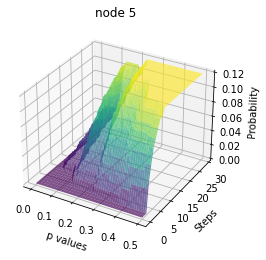}
    \end{subfigure}
	\caption{Probability distribution of the walker at all nodes of the graph $T_6$ under NMD noise. The channel parameters are the same as for path graph.}
	\label{Irregular_prob_NMD}
\end{figure}

	Figure \ref{Irregular_prob_NMD} depicts the probability distributions of the walker at all vertices of $T_6$ under NMD noise. At node $1$, the probability decreases while at other nodes it increases and becomes stationary after a few steps of quantum walk.

\begin{figure}
		\centering
		\includegraphics[scale = .5]{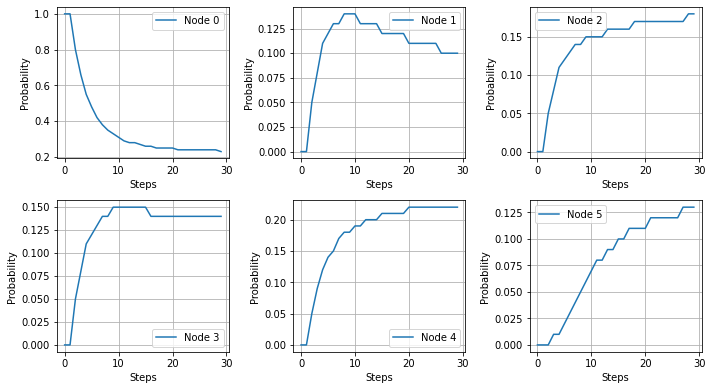}
		\caption{Probability distribution of the walker at all nodes of the graph $T_6$ under the non-Markovian depolarization noise. The channel parameters are the same as for path graph.}
		\label{Irregular_prob_depolarizaton}
	\end{figure}

	The coin operators for the graph $T_{6}$ under a non-Markovian depolarization channel are defined by	
	\begin{equation}
	\begin{split}
	&C_{(u,u)} = \sqrt{1 + \frac{4(1-p)\Lambda_1}{5}}~ U_{(u,u)}, ~\text{for}~ u = 0 ~\text{and} ~ 4;  \\
	&C_{(1,1)} = \sqrt{1 + \frac{(1-p)\Lambda_1}{2}}~ U_{(1,1)}; \\ 
	&C_{(2,2)} = \sqrt{1 + \frac{3(1-p)\Lambda_1}{4}}~ U_{(2,2)}; \\
	&C_{(u,u)} = \sqrt{1 + \frac{2(1-p)\Lambda_1}{3}}~ U_{(u,u)}, ~\text{for}~ u = 3 ~\text{and} ~ 5; \\
	&C_{(0,v)} = \sqrt{\frac{p\Lambda_2}{5}}~ U_{(u,v)}, ~\text{for}~ 1 \leq  v \leq (n - 2); \\
	&C_{(1,0)} = \sqrt{\frac{p\Lambda_2}{2}}~ U_{(1,0)}; \\
	&C_{(2,v)} = \sqrt{\frac{p\Lambda_2}{4}}~ U_{(2,v)}; \\
	&C_{(u,v)} = \sqrt{\frac{p\Lambda_2}{3}}~ U_{(u,v)}, ~\text{for}~ u = 3 ~\text{and} ~ 5; \\
	&C_{(4,v)} = \sqrt{\frac{p\Lambda_2}{5}}~ U_{(4,v)}.
	\end{split}  
	\end{equation}
	The probability distribution of the walker on the graph $T_{6}$ under non-Markovian depolarization noise is depicted in Figure \ref{Irregular_prob_depolarizaton}. The probability of the walker decreases at vertex $0$ while it increases at the other nodes with time-steps and becomes stationary after some time-steps.
	
	\section{Conclusion}
	
	We focus on the effect of noise in discrete-time quantum walks on arbitrary graphs, going beyond the case of regular graphs. A new model is proposed to investigate the effect of noise in the Discrete-Time Open Quantum Walk (DTOQW). To study the impact of noise in discrete-time quantum walks, the coin and noise operators are usually applied separately. In contrast, in this work, we replace the coin operators with the Kraus operators of the corresponding noisy channels. Weyl operators are used to develop the Kraus operators in arbitrary dimensions.	
	
	Noise from the surrounding environment inevitably affects a quantum walk. We have taken into consideration three physically relevant noise models: non-Markovian amplitude damping, non-Markovian dephasing, and non-Markovian depolarization. The Kraus operators corresponding to each considered non-Markovian noise are generalized for the qudit system to make them compatible for application on an arbitrary graph. We consider a few specific examples of graphs for illustration which includes the path graph, cyclic graph, star graph, complete graph, completely bipartite graph and an arbitrary graph. The probability distribution of the walker at different vertices of the considered graph is calculated. This shed light on the type of graph on which the DTOQW was implemented. Further, the fidelity between the two states of the walker at different time steps and the coherence of the state were also studied. We observed that under the considered non-Markovian noise channels, the coherence and fidelity patterns are qualitatively similar for all the considered graphs bringing out their dependence on the nature of noise rather than the structure of the graphs.	
	
	This article provides a well-established theoretical framework for a new DTOQW model. In this model, the walker can move on an arbitrary graph. Networks are indispensable in the real world. Graph theory builds up the mathematical foundation of all networks, including social, biological, and computer networks. Many network-related algorithms use the classical random walks on arbitrary graphs, such as the PageRank algorithm. The new quantum walk model was utilized to create a quantized version of the classical PageRank algorithm \cite{dutta2024discrete}. Similarly, we can use the new quantum walk to detect communities and clusters of vertices, as well as find other centrality measures on the network. Therefore, DTOQW would aid in the modeling of network behavior. In this setting, our findings would be useful.

	\section*{Funding}
	
	This work is partially supported by the project ``Transmission of quantum information using perfect state transfer" (CRG/2021/001834) funded by the Science and Engineering Research Board, Government of India. 
	
	\appendix
	\section{Appendix: Completeness condition for Kraus operators}\label{Completeness_condition}
	\label{appendix}
	
	Consider the cion operators mentioned in equation (\ref{ADC}). Now,
	\begin{equation}
		\begin{split}
		& C_{(u, u)}^\dagger C_{(u, u)} + \sum_{\overrightarrow{(u, v)} \in E(\overrightarrow{G})} C_{(u, v)}^\dagger C_{(u, v)} \\
		= & \left[\ket{u}\bra{u} + \sqrt{1 - \lambda(t)} \sum_{\overrightarrow{(u, v)} \in E(\overrightarrow{G})} \ket{v}\bra{v} + \sum_{\overrightarrow{(u, w)} \notin 	E(\overrightarrow{G})} \ket{w}\bra{w}\right]^\dagger \\
		& \hspace{.1cm} \times \left[\ket{u}\bra{u} + \sqrt{1 - \lambda(t)} \sum_{\overrightarrow{(u, v)} \in E(\overrightarrow{G})} \ket{v}\bra{v} + \sum_{\overrightarrow{(u, w)} \notin E(\overrightarrow{G})} \ket{w}\bra{w}\right] + \sum_{\overrightarrow{(u, v)} \in E(\overrightarrow{G})} \left[\sqrt{\lambda(t)} \ket{u}\bra{v}\right]^\dagger \left[\sqrt{\lambda(t)} \ket{u}\bra{v}\right]\\
		= & \ket{u}\braket{u | u}\bra{u} + (1 - \lambda(t)) \sum_{\overrightarrow{(u, v)} \in E(\overrightarrow{G})} \ket{v}\braket{v | v}\bra{v} + \sum_{\overrightarrow{(u, w)} 	\notin E(\overrightarrow{G})} \ket{w}\braket{w | w}\bra{w} + \lambda(t) \sum_{\overrightarrow{(u, v)} \in E(\overrightarrow{G})} \ket{v}\braket{u | v}\bra{v} \\
		= & \ket{u}\bra{u} + (1 - \lambda(t)) \sum_{\overrightarrow{(u, v)} \in E(\overrightarrow{G})} \ket{v}\bra{v} + \sum_{\overrightarrow{(u, w)} \notin E(\overrightarrow{G})} 	\ket{w}\bra{w} + \lambda(t) \sum_{\overrightarrow{(u, v)} \in E(\overrightarrow{G})} \ket{v}\bra{v} \\
		= & \ket{u}\bra{u} + \sum_{\overrightarrow{(u, v)} \in E(\overrightarrow{G})} \ket{v}\bra{v} + \sum_{\overrightarrow{(u, w)} \notin E(\overrightarrow{G})} \ket{w}\bra{w} = I_n.
		\end{split}
	\end{equation}.
		Thus, the Kraus operators modelling the DTOQW evolution under the non-Markovian ADC satisfies the completeness criterion.
	
		Now, consider the coin operators, defined in equation (\ref{NMD}), for the DTOQW  under non-Markovian depolarization channel whose completeness condition is shown below. 
		\begin{equation}
			\begin{split}
				& C_{(u, u)}^\dagger C_{(u, u)} + \sum_{\overrightarrow{(u, v)} \in E(\overrightarrow{G})}  C_{(u, v)}^\dagger C_{(u, v)} \\
				= & \left[\sqrt{1 - \kappa(p)}U_{0,0}\right]^\dagger\left[\sqrt{1 - \kappa(p)}U_{0,0}\right] +  \sum_{\overrightarrow{(u, v)} \in E(\overrightarrow{G})}  	\left[\sqrt{\frac{\kappa(p)}{d_u}} U_{u,v}\right]^\dagger \left[\sqrt{\frac{\kappa(p)}{d_u}} U_{u,v}\right]\\
				= & (1-\kappa(p))U_{0,0}^\dagger U_{0,0} + \frac{\kappa(p)}{d_u} \sum_{\overrightarrow{(u, v)} \in E(\overrightarrow{G})} U_{u,v}^\dagger U_{u,v} = (1  -\kappa(p))I_{n} + \frac{\kappa(p)}{d_u} d_u I_{n} = I_{n}.
			\end{split}
		\end{equation}
	
		The corresponding scenario for the DTOQW under the influence of the non-Markovian depolarization channel, is described below, where the coin operators are as in equation (\ref{Depolarization}).
		\begin{equation}
		\begin{split}
		& C_{(u, u)}^\dagger C_{(u, u)} + \sum_{\overrightarrow{(u, v)} \in E(\overrightarrow{G})} C_{(u, v)}^\dagger C_{(u, v)} \\
		=& \left[\sqrt{1 + \frac{d_u(1-p)\Lambda_1}{d_u+1}} U_{0,0}\right]^\dagger\left[\sqrt{1 + \frac{d_u(1-p)\Lambda_1}{d_u+1}} U_{0,0}\right] + \sum_{\overrightarrow{(u, v)} \in E(\overrightarrow{G})} \left[\sqrt{\frac{p\Lambda_2}{d_u+1}}U_{u, v}\right]^\dagger\left[\sqrt{\frac{p\Lambda_2}{d_u+1}}U_{u,v}\right]\\
		= & \left(1 + \frac{d_u(1-p)\Lambda_1}{(d_u+1)}\right) U_{0,0}^\dagger U_{0,0} + \frac{p\Lambda_2}{(d_u+1)}\sum_{\overrightarrow{(u, v)} \in E(\overrightarrow{G})} U_{u,v}^\dagger U_{u,v}\\
		= & \left(1+\frac{d_u(1-p)\Lambda_1}{d_u+1}\right)I_{n} + \left(\frac{p\Lambda_2}{(d_u+1)}\right)(d_u) I_{n}\\
		= & \left(1 + \frac{d_u(1-p)\Lambda_1}{(d_u+1)} + \frac{d_u p\Lambda_2}{d_u+1}\right)I_{n} = (1 + \frac{d_u}{d_u+1}\left((1-p)\Lambda_1 + d_u\Lambda_2)\right)I_{n} = I_{n}.
		\end{split}
		\end{equation}

\end{document}